\def\lam{\mbox{$\:\lambda\ $}}
\newcommand\lamlam{\mbox{$\:\lambda\lambda $ }}
\def\ha{{H$\alpha$}}
\def\hb{{H$\beta$}}

\def\kms{\:\rm{\,km\,s^{-1}}}

\def\LUM{\:{\rm erg\:s^{-1}}}
\def\FLUX{\:{\rm erg\:cm^{-2}\:s^{-1}}}

\def\VEL{\:{\rm km\:s^{-1}}}

\def\etal{{\it et\:al.}}

\def\oiiL{[\ion{O}{2}] $\lambda 3727$}

\def\sii{[\ion{S}{2}]}
\def\nii{[\ion{N}{2}]}
\def\oi{[\ion{O}{1}]}

\def\oiii{[\ion{O}{3}]}
\def\feii{[\ion{Fe}{2}]}

\def\hii{\ion{H}{2}}
\def\hi{\ion{H}{1}}

\documentclass[12pt,preprint]{aastex}
\usepackage{relsize}

\begin{document}


\newcommand{\MSOL}{\mbox{$\:M_{\sun}$}}  

\newcommand{\EXPN}[2]{\mbox{$#1\times 10^{#2}$}}
\newcommand{\EXPU}[3]{\mbox{\rm $#1 \times 10^{#2} \rm\:#3$}}  
\newcommand{\POW}[2]{\mbox{$\rm10^{#1}\rm\:#2$}}
\newcommand{\SING}[2]{#1$\thinspace \lambda $#2}
\newcommand{\MULT}[2]{#1$\thinspace \lambda \lambda $#2}
\newcommand{\CHINU}{\mbox{$\chi_{\nu}^2$}}
\newcommand{\vsini}{\mbox{$v\:\sin{(i)}$}}
\newcommand{\LSOL}{\mbox{$\:L_{\sun}$}}

\newcommand{\fuse}{{\it FUSE}}
\newcommand{\hst}{{\it HST}}
\newcommand{\iue}{{\it IUE}}
\newcommand{\euve}{{\it EUVE}}
\newcommand{\einstein}{{\it Einstein}}
\newcommand{\rosat}{{\it ROSAT}}
\newcommand{\chandra}{{\it Chandra}}
\newcommand{\xmm}{{\it XMM-Newton}}
\newcommand{\swift}{{\it Swift}}
\newcommand{\asca}{{\it ASCA}}
\newcommand{\galex}{{\it GALEX}}
\newcommand{\cxo}{CXO}

\slugcomment{Accepted for publication in {\em The Astrophysical Journal}}

\shorttitle{Spectroscopy of Supernova Remnants in M83}
\shortauthors{Winkler \etal}

\title{
A Spectroscopic Study of the Rich Supernova Remnant Population in M83\footnote{Based on observations obtained at the Gemini Observatory, which is operated by the Association of Universities for Research in Astronomy, Inc., under a cooperative agreement with the NSF on behalf of the Gemini partnership: the National Science Foundation (United States), the National Research Council (Canada), CONICYT (Chile), Ministerio de Ciencia, Tecnolog\'{i}a e Innovaci\'{o}n Productiva (Argentina), and Minist\'{e}rio da Ci\^{e}ncia, Tecnologia e Inova\c{c}\~{a}o (Brazil).}}

\author{
P. Frank Winkler\altaffilmark{1,4}
William P. Blair\altaffilmark{2,4},
 and
Knox S. Long\altaffilmark{3,5}
}
\altaffiltext{1}{Department of Physics, Middlebury College, Middlebury, VT, 05753; 
winkler@middlebury.edu}
\altaffiltext{2}{The Henry A. Rowland Department of Physics and Astronomy, 
Johns Hopkins University, 3400 N. Charles Street, Baltimore, MD, 21218; 
wpb@pha.jhu.edu}
\altaffiltext{3}{Space Telescope Science Institute, 3700 San Martin Drive, 
Baltimore, MD, 21218;  long@stsci.edu}
\altaffiltext{4}{Visiting Astronomer, Gemini South Observatory, La Serena, Chile}
\altaffiltext{5}{Eureka Scientific, Inc.
2452 Delmer Street, Suite 100,
Oakland, CA 94602-3017}

\begin{abstract}
We report the results from a spectrophotometric study sampling the $\gtrsim 300$ candidate supernova remnants (SNRs) in M83 identified through optical imaging with  Magellan/IMACS and {\em HST}/WFC3\@.  Of the 118 candidates identified based on a high \sii\,\lamlam 6716,6731 to \ha\ emission ratio, 117 show  spectroscopic signatures of shock-heated gas, confirming them as SNRs---the largest uniform set of SNR spectra for any galaxy.  Spectra of 22 objects with a high \oiii\,\lam 5007 to \ha\ emission ratio, selected in an attempt to identify young ejecta-dominated SNRs like Cas A, reveal only one (previously reported) object   with the broad ($\gtrsim 1000 \kms$) emission lines characteristic of ejecta-dominated SNRs, beyond the known  SN1957D remnant.  The other 20 \oiii-selected candidates include planetary nebulae, compact \hii\ regions, and one background QSO\@.  Although our spectroscopic sample includes 22 SNRs  smaller than 11 pc, none of the other objects shows broad emission lines; 
instead their spectra stem from relatively slow ($\sim200 \kms$) radiative shocks propagating into the metal-rich interstellar medium of M83.   
With six SNe in the past century, one might expect more of M83's small-diameter SNRs to show evidence of ejecta; this appears not to be the case. We attribute their absence to several factors, including that  SNRs expanding into a dense medium evolve quickly to the ISM-dominated phase, and that SNRs  expanding into  regions already evacuated by earlier SNe are probably very faint.  

Subject Headings: galaxies: individual (M83) -- galaxies: ISM  -- ISM: supernova remnants

\end{abstract}

\section{Introduction \label{sec_intro}}

No problem is more central to astrophysics than understanding how stars are born, live, and die.  This cycle is responsible for enriching the cosmos in heavy elements from the Big Bang to the present and largely determines the luminosity, spectral energy distribution, and morphology of galaxies over cosmic time.  Supernova remnants (SNRs), which represent the ashes from one generation of stars and provide the raw material for another, provide an important window through which we can view the stellar cycle.  For investigating the overall SNR population, we must look to nearby galaxies, where SNRs can be readily studied free from the foreground extinction that plagues most of the SNRs in the Milky Way, and where all the objects are effectively at the same distance.  

M83 (NGC\,5236) is a classic grand-design SAB(s)c spiral galaxy with a starburst nucleus, active star formation along the arms, and prominent dust lanes \citep{elmegreen98}.  It has played host to six recorded supernovae (SNe) in the past century, ranking behind  only  NGC\,6946 (with nine) and the far more distant NGC\,3690 and NGC\,4303 (seven each) \citep{barbon99}.   At a distance of 4.61 Mpc \citep{saha06}, M83 is close enough to be studied effectively with current generations of telescopes (1\arcsec\,=\,22 pc), and it is nearly face-on.  Hence, M83  arguably provides the most comprehensive view of any galaxy where such active star formation and destruction are taking place.   The integrated effects of this active star formation process are manifest through the generally high metallicity and the chemical abundance gradients measured by spectroscopy of \hii\ regions across the $\sim$10\arcmin\ diameter bright optical disk  \citep{{bresolin02}, {pilyugin06a}, {pilyugin10}, {bresolin16}}.  A fainter and much more extended disk is seen in \hi\  and in GALEX ultraviolet imaging data \citep{{huchtmeier81}, {thilker05}, {bigiel10}}.


Five of the six historical SNe in M83 have types of either Ib, II, or IIP, all of which result from core-collapse of massive stars \citep{{barbon99}, {williams15}}.  Simple extrapolation from the recent past thus leads us to expect that there must have been dozens of core-collapse SNe in M83 within the past millennium.  In addition, there should be  many more older SNRs, since they typically remain visible for tens of thousands of years, depending on local conditions in the interstellar medium (ISM) around each object.  These considerations have led us and collaborators to undertake extensive multi-wavelength studies of M83, one important goal of which has been to identify  SNRs in M83 and to characterize them---both individually and as a population.

Imaging studies by ourselves and others have identified numerous supernova remnants (SNRs) in M83.  The most commonly used technique for optically identifying SNR candidates is to find nebulae with strong \sii\,\lamlam 6716,6731 emission relative to \ha\ in digital images.  This criterion has long proved effective for discriminating SNRs, especially evolved ones dominated by matter swept up from the interstellar medium, from other types of nebulae.  Its physical basis stems from the fact that the supernova blast wave rapidly heats and ionizes material entering the shock; this is followed by gradual cooling and recombination.  Emission from this cooling tail is characterized by forbidden lines from a range of ionization states, including S$^+$.  
By contrast, in photo-ionized regions radiation from embedded hot stars maintains  a higher ionization state; sulfur exists mainly as S$^{++}$, and the \sii\ emission lines are relatively weak.  SNRs typically have \sii/\ha\ ratios $\gtrsim 0.4$, while \hii\ regions typically have \sii/\ha$\: \lesssim 0.2$ \citep[e.g.][]{mathewson72,dodorico80,levenson95,blair97,matonick97}.  

The first search for SNRs in M83 was carried out by \citet[][hereinafter BL04]{blair04}.  They used the \sii/\ha\ line ratio in CCD images to identify 71 emission nebulae as probable SNRs in M83, and then conducted follow-up longslit spectroscopy of 25 of these candidates, confirming 23.  BL04 also carried out a separate search for \oiii-bright nebulae in order to identify ejecta-dominated SNRs, similar to Cas A \citep[e.g.,][]{kirshner77, fesen01} 
in our Galaxy or 1E 0102-7219 in the Small Magellanic Cloud \citep[e.g.,][]{dopita84, blair00}.  
Such young remnants as these in M83 would likely be unresolved at ground-based resolution (e.g., at the distance of M83, Cas A would have a diameter of only 0\farcs 15) and could be confused with planetary nebulae.  Extended  nebulae with  high ratios of \oiii\lam 5007 to \ha\  are likely to be normal ISM-dominated SNRs with high enough shock velocities to excite \oiii, or perhaps nebulae excited by early-type Wolf-Rayet stars.  BL04 noted 36 nebulae with an \oiii/\ha\ ratio between 0.25 and 0.80, almost half of which were also on their list of SNR candidates with high \sii/\ha\ ratios.  The only bona fide ejecta-dominated SNR confirmed as part of this survey was the remnant of the historical SN\,1957D, recovered as a faint, unresolved \oiii\ nebula at the SN position \citep{long89}.

\citet[hereinafter D10]{dopita10}  reported the results from an imaging study, carried out  with the  Wide Field Camera 3 (WFC3) on the {\it Hubble} Space Telescope, of a single $162\arcsec \times 162\arcsec$ field in M83 that included the complex nuclear region and part of one spiral arm.  They identified 60 SNR candidates that are relatively bright in both \sii\ and \oiiL\ relative to \ha, only 12 of which had been identified by BL04.  In addition, D10  identified six (slightly) extended nebulae with \oiii\ emission that they suggested might be young, ejecta-dominated SNRs, one compact \oiii\ nebula with a corresponding X-ray source that they regarded as a very strong candidate to be an ejecta-dominated SNR, and the possible counterpart to SN1968L deep within the complex starburst nuclear region.

Subsequently, \citet [][hereinafter B12]{blair12} reported on a comprehensive search for SNRs in M83  from emission-line imaging of the entire bright optical disk of M83 using the 6.5m Magellan-I telescope and the IMACS instrument under conditions of exceptional seeing ($\lesssim 0\farcs 5$).   They found a total of 225 SNR candidates with \sii/\ha\ line ratios $> 0.4$\@, including all but  three of the faintest  candidates from BL04.  

B12 also carried out a search based on the \oiii\,\lam 5007:\ha\  ratio, similar to that of BL04 but with better sensitivity and angular resolution, that led to the identification of 46 additional \oiii-selected objects.  In order to weed out planetary nebulae, B12 required that these objects have either (1) associated X-ray emission, based on the the {\em Chandra} ACIS survey by \citet[][hereinafter SW03]{soria03} or a preliminary analysis of data from our much deeper {\em Chandra} ACIS survey \citep[hereinafter L14]{long14}; and/or (2) a spatial extent large enough to be resolved (size $\gtrsim 0\farcs 6 \approx 13$ pc).  

Most recently, \citet[hereinafter B14]{blair14}  reported results from a survey of M83 from {\em HST}, much more extensive that that of D10 and comprising seven WFC3 fields.  In that paper, we focused narrowly on one topic: the population of {\em young}\ SNRs---those whose sizes (mostly $< 0\farcs 5$) make them difficult to fully characterize from the ground.  Analysis of both optical and IR (\feii\ 1.644 $\mu$m) emission-line images led to a list of 63 candidates, of which 37 had previously been included in the B12 and/or D10 catalogs.  Further analysis of the {\em Chandra} ACIS data led to identification of 26 of the 63 objects as soft X-ray sources, a further indicator of youthful vigor in SNRs.   

Here we describe the results of a spectroscopic study of dozens of SNR candidates in M83, carried out using the Gemini Multi-Object Spectrograph (GMOS) on the 8.2m Gemini South telescope in April 2011, plus additional GMOS spectra from April-June 2015.  
We have previously described the spectra from two objects of singular interest:  SN\,1957D,  in \citet{long12} where we reported the discovery of X-ray emission from this remnant of the historical SN; and a second apparently very young SNR that we have designated B12-174a  \citep[close neighbor to object 174 in the B12 catalog,][]{blair15}.  B12-174a must be under 100 years old, based on its small size and very broad emission lines that indicate high-velocity ejecta, but no historical evidence of its explosion has so far been discovered.

The present paper is organized as follows: Section 2 reports the target selection, observations, and data reduction.  Section 3 summarizes our results---line fluxes and other data for 140 SNR candidates.  Section 4 provides an analysis of the observations and a brief discussion, and a summary follows in Section 5.

\section{Observations and Data Reduction \label{sec_obs}}

We used the Gemini Multi-Object Spectrograph (GMOS) on the 8.2m Gemini-South telescope to obtain all the spectra reported here.  Most were obtained in a classically scheduled observing run  on 2011 April 7-9 (UT).
In advance of this run, we designed seven custom masks, each with 20-30 slitlets, whose positions we determined from our 2009 Magellan IMACS images, together with  short $R$-band pre-images of several M83 fields taken with GMOS earlier in 2011 as part of the spectroscopy program.  
We selected objects from the lists of SNR candidates in D10 and in a preliminary version of the catalog that later appeared in B12.
 Slitlets in one or more of our seven masks were placed on 107 distinct SNR candidates, including ones with a range of sizes, galactocentric radii, and ISM environments (locations in arms and in 
 inter-arm regions).\footnote{Exceptions are the very outermost regions of the galaxy, where the sparse population would have made less efficient use of the $5\farcm 5 \times 5\farcm 5$ GMOS field, and the innermost nuclear region, where source confusion and the high star density would have made sky subtraction difficult.}  
 
\begin{figure}
\epsscale{0.95}
\plotone{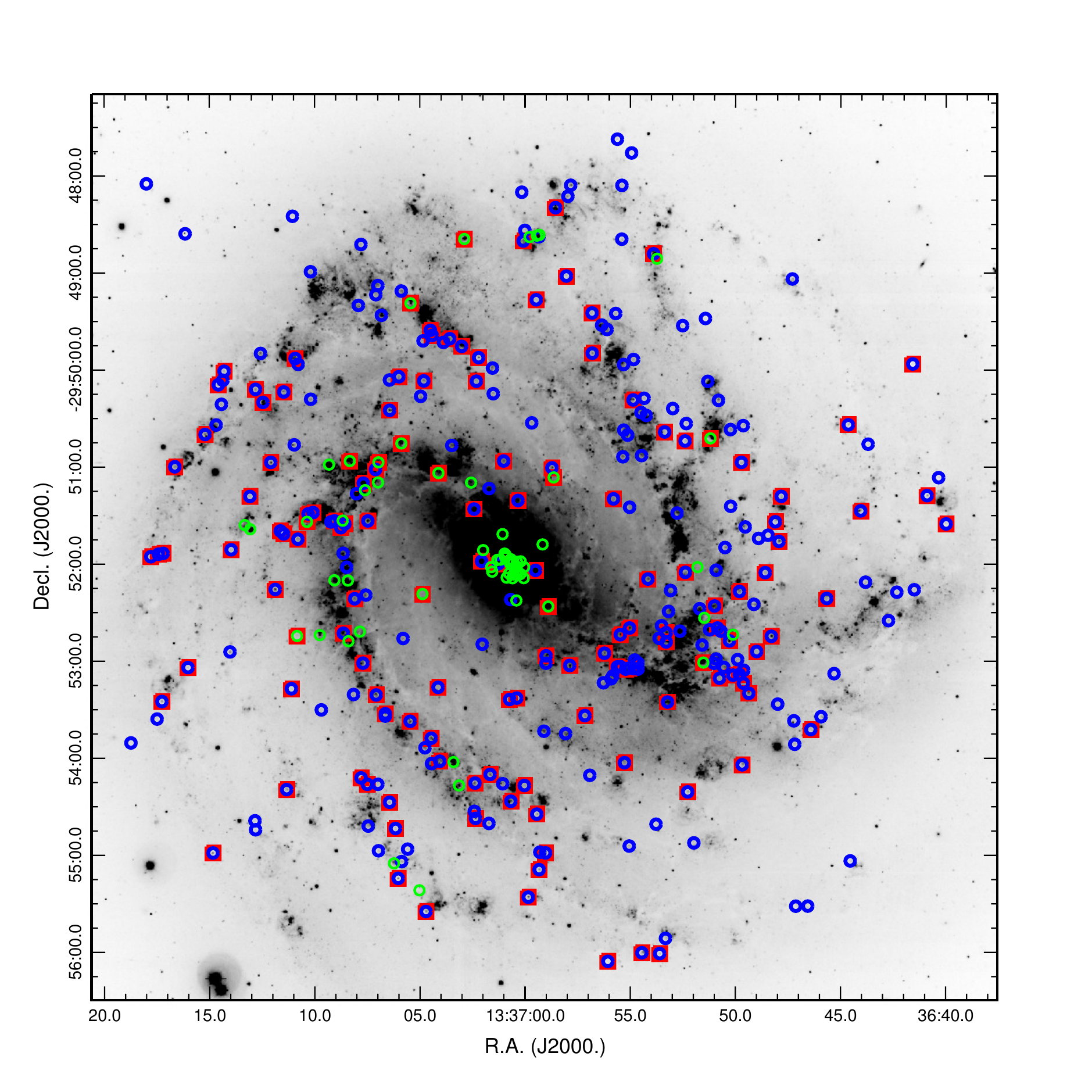}
\caption{An \ha\ image of M83 taken with IMACS on the Magellan II telescope in 2009 \citep[for details, see][]{blair12}.  All the SNR candidates from \citet{blair12} are marked with blue circles; additional (non-duplicate) candidates from \citet{dopita10} by green  circles.  The red squares indicate objects for which we obtained GMOS spectra (listed in Table 2).  Note that the objects with spectra are well distributed around the galaxy, with the exception of the very outermost regions.  We chose not to target those areas because the sparse population would have made less efficient use of the GMOS capabilities.  The overall field shown is 9.3\arcmin\ square, oriented north up, east left.}
\label{fig.galaxy}
\end{figure}

We later obtained spectra for two additional masks (which we refer to as masks 8 and 9 for simplicity) in a queue-scheduled program (GS-2015A-Q-90) during the 2015A semester.   For this program we concentrated especially on small-diameter objects, many of which we had by then identified in the \hst/WFC3 images (D10, B14), though we also included several additional objects from the B12 lists.  In the 2011 and 2015 runs together we obtained spectra of 140 different SNR candidates.
Figure \ref{fig.galaxy} shows all the objects from the B12, D10, and B14 catalogs, and highlights those for which we obtained spectra (red boxes).  In addition to the SNR candidates, we also placed a number of slitlets on prominent \hii\ regions and bright planetary nebulae for comparison purposes.

For all the spectra we used GMOS-S with the 600 lines\,mm$^{-1}$ grating designated G5323 and a GG455 cut-off filter to block second-order spectra.   The detector in 2011 was a mosaic of e2v CCD chips, binned $\times 2$ in the spatial direction (for a scale of 0\farcs 146 pixel$^{-1}$) and $\times 4$ in the dispersion direction.  
The dispersion was 1.84 \AA\,  pixel$^{-1}$ (binned), resulting in coverage of the spectral range of at least \hb\ through \sii\lamlam 6716,\,6731 for virtually all the objects.\footnote{The wavelength coverage naturally varied with slitlet position in the dispersion direction.}  Our  masks had  slitlet  widths from 1\farcs 25 to 1\farcs 75, with wider slits used for the larger objects, and lengths of 6\arcsec\ or longer to permit local background sky subtraction.   
In 2011, at least three frames at different wavelength settings were taken with each mask, in order to eliminate cosmic rays and to cover wavelength gaps between chips.  For calibration, we took quartz flats and CuAr arc frames  immediately before or after the science exposure(s) with each mask and wavelength setting, as well as longslit spectra of a number of spectrophotometric standard stars from the list of \citet{hamuy92}.  

During the 2015A semester, the e2v CCDs in GMOS-S had  been replaced with a mosaic of Hamamatsu CCDs.  Unfortunately, one of these chips soon developed a problem with one of the amplifiers that resulted in data loss from a portion of the detector. To work around the amplifier problem, we used five widely spaced wavelength settings in order to bridge over sections of the detector where data were missing, and took a set of three frames at each setting to eliminate cosmic rays. 
We again used 2 (spatial) $\times 4$ (dispersion) binning; both the spatial scale and dispersion were essentially identical to that of the earlier chips.  A journal of the science observations from both 2011 and 2015 appears in Table \ref{obs_log}.

The data were processed using standard procedures for bias subtraction, flat-fielding, wavelength calibration, and flux calibration from the {\tt gemini} package in IRAF.\footnote{IRAF is distributed by the National Optical Astronomy Observatory, which is operated by the Association of Universities for Research in Astronomy, Inc., under cooperative agreement with the National Science Foundation.}  During the processing, the 2-D spectra from different slitlets were separated; we examined each of these individually to determine the optimum background sky subtraction region.  Many of the objects are located in regions with bright surrounding galactic background (both continuum and emission lines) from M83, so the ability to subtract a representative  local background in the immediate vicinity of each object is important for obtaining accurate spectra.  Finally, we extracted one-dimensional spectra by summing rows containing each object using an algorithm that allows for a small linear slope  from the blue to the red end of each spectrum.

\begin{figure}
\epsscale{1.0}
\plotone{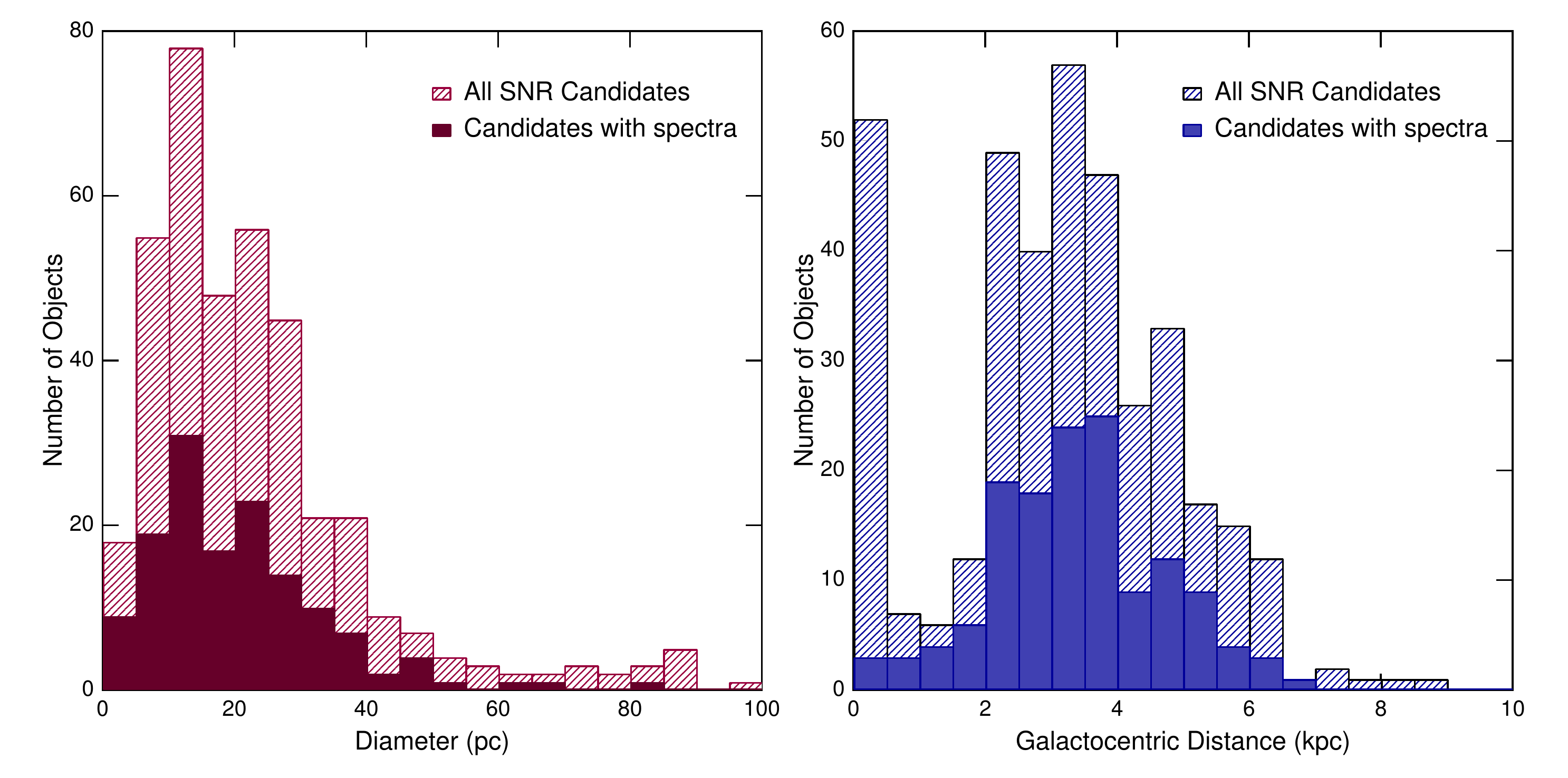}
\caption{({\em a, left}) Histogram of SNR diameter for all candidates and for those with spectra. ({\em b, right}) Histogram of galactocentric distance (GCD) for all candidates and for those with spectra.  Our sample is representative of the entire ensemble of sizes and distances, except for the smallest and largest GCDs.  As noted in the text, we targeted few objects in the nuclear region of M83 because of confusion and the difficulty of background subtraction, and none in the very outermost regions because this would have been an inefficient use of the GMOS. }
\label{fig_histograms}
\end{figure}

Tables \ref{table-SNR} and \ref{table-O3} list all the SNR candidates for which we obtained spectra.  In both tables, column 1 gives the object name from  the B12, B14, and/or D10 lists; column 2 provides some alternative names for the objects, columns 3 and 4 give the position; column 5 the diameter (assuming a distance to M83 of 4.61 Mpc).  Most of the objects lie within the footprint of the WFC3 observations, and for the great majority of these we measured the diameters from the WFC3 images.  For candidates in the outer galaxy, plus a few low-surface-brightness ones within the WFC3 footprint, we used the Magellan images instead, adjusted for seeing.   Column 6 gives the galactocentric distance; and column 7 notes objects that were detected in the {\em Chandra} ACIS X-ray survey by \citet{long14}.  Column 8 gives the mask and slitlet number used for extracting the one-dimensional spectra.  Several objects were observed with more than one mask; listed here is the one with the highest signal-to-noise ratio.   As an indication that the sample with spectra is representative of the overall population of SNR candidates in M83, we show  histograms of number vs.\ diameter in Fig.~\ref{fig_histograms}a and of number vs.\ galactocentric distance in Fig.~\ref{fig_histograms}b. The singular exception is that we purposely ignored the crowded central starburst region since spectra in this region would have been too confused.


\section{Results}

Among all nine masks, we have obtained spectra for 140 possible SNRs: 127 of the 271 SNRs and candidates cataloged by B12 (108 of these \sii-selected, their Table 2; 19 \oiii-selected, their Table 3), plus seven  candidates from D10 (five \sii-selected, their Table 2; two \oiii-selected, their Table 4) that are not duplicated in the B12 list, plus six additional objects identified in the full WFC3 survey (B14).
Taking all the objects together, whether previously confirmed or not and whether identified from Magellan or WFC3 data, there are a total of 118 \sii-selected (probably ISM-dominated) and 22 \oiii-selected objects.  In this last grouping, we have included the young ejecta-dominated object B12-174a (which met criteria for selection based on both its \sii\ and \oiii\ lines) and the remnant of SN1957d with the \oiii-selected group for the purposes of the subsequent discussion.

\begin{figure}
\begin{centering}
\includegraphics[scale=0.58]{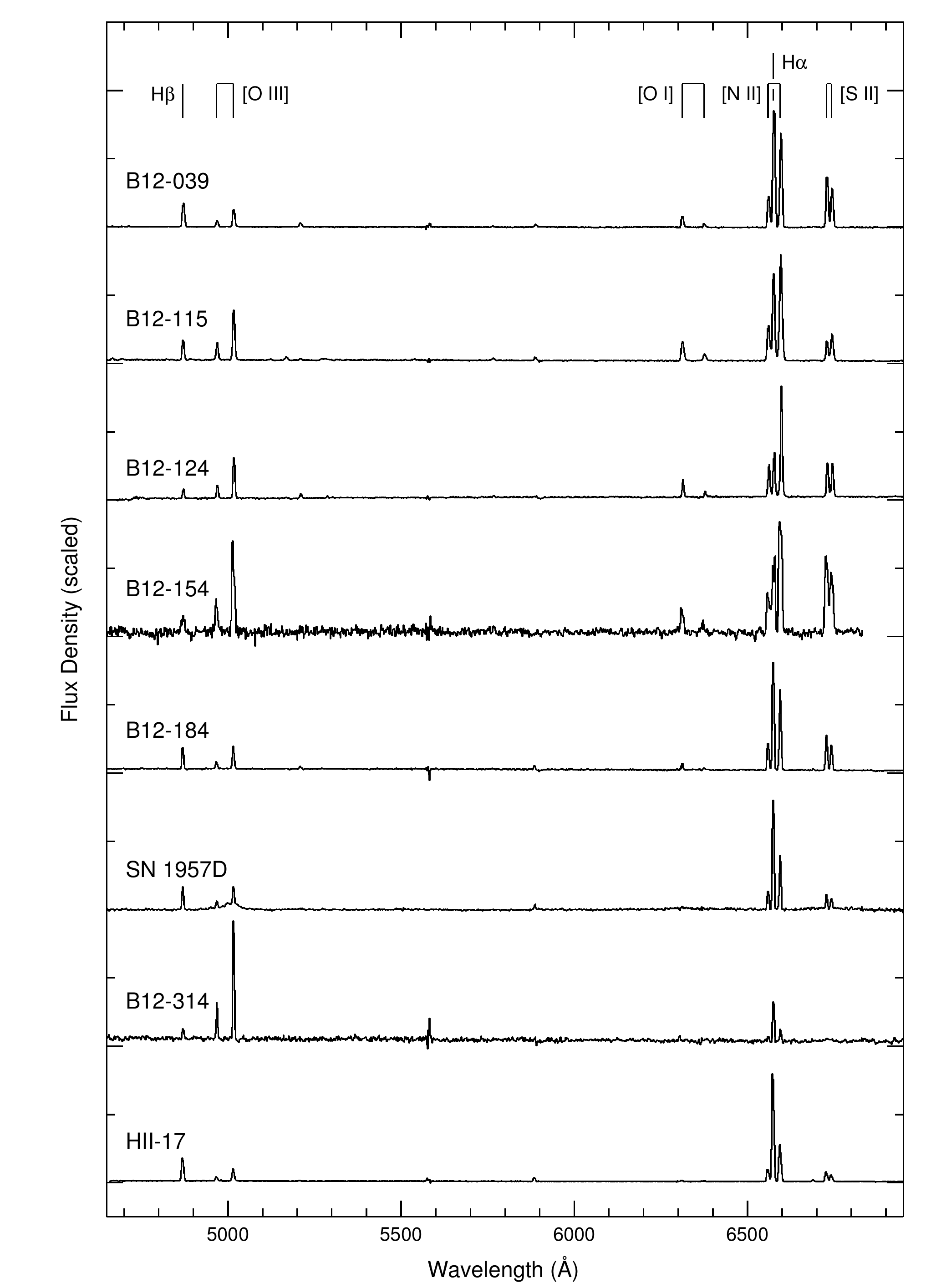}
\caption{Several typical examples of one-dimensional spectra extracted from the two-dimensional GMOS data.  The top five traces are all \sii-selected candidates, confirmed as SNRs through these spectra. The SN1957D spectrum includes a near-coincident \hii\ region with narrow-lines; the broad oxygen and sulfur lines are less apparent here than in the version with a magnified vertical scale in \citet[][Fig.~4]{long12}. B12-314 is an \oiii-selected candidate, and is unresolved even with {\em HST}; it is probably a bright planetary nebula.  The X-ray source L14-135 is located $< 1\arcsec$ away, but this may be a chance coincidence with an X-ray binary. The bottom trace is a typical \hii\ region.  The spectra have been arbitrarily scaled and vertically offset for clarity.}
\label{fig_multiple_spectra}
\end{centering}
\end{figure}

Since the regions covered by different masks overlap considerably, several of the objects were observed with  more than one mask. In cases with multiple spectra of the same object, we present here the results for the one with the best signal-to-noise (where one is clearly superior), or have combined the multiple spectra (where two or more have comparable signal-to-noise).  In Fig.~\ref{fig_multiple_spectra} we show several typical examples of our background-subtracted and extracted one-dimensional spectra.

For all the 1-D spectra, we performed Gaussian fits to the prominent emission lines---\hb, \oiii\,\lamlam 4959, 5007, \oi\lamlam 6300, 6363, \nii\lamlam\,6548, 6583, \ha, and \sii\lamlam 6716, 6731---where we fit the central wavelength, integrated flux (relative to the local continuum), and FWHM for each line.  We give the fluxes obtained for most of these lines (not including \oiii\ \lam 4959, \oi \lam 6363, or \nii \lam 6548, all of whose fluxes relative to the stronger member of their respective doublets are determined by atomic physics) in Table \ref{table-snr-flux} for the \sii-selected objects and Table \ref{table-O3-flux}  for the [O III]-selected objects.   

We do not quote uncertainties for the line fluxes in Tables \ref{table-snr-flux} and \ref{table-O3-flux}. Although our fitting routine does give a formal error, in most cases the actual uncertainty is limited by systematic errors in subtracting the night sky and galaxy background in our two-dimensional spectra from slitlets that are typically only 6\arcsec-10\arcsec\ long.  Such effects are most significant for lines that are present in the diffuse night sky and/or in much of the M83 disk: \ha, \hb, and \oi \lamlam 6300, 6363.  

\begin{figure}
\plotone{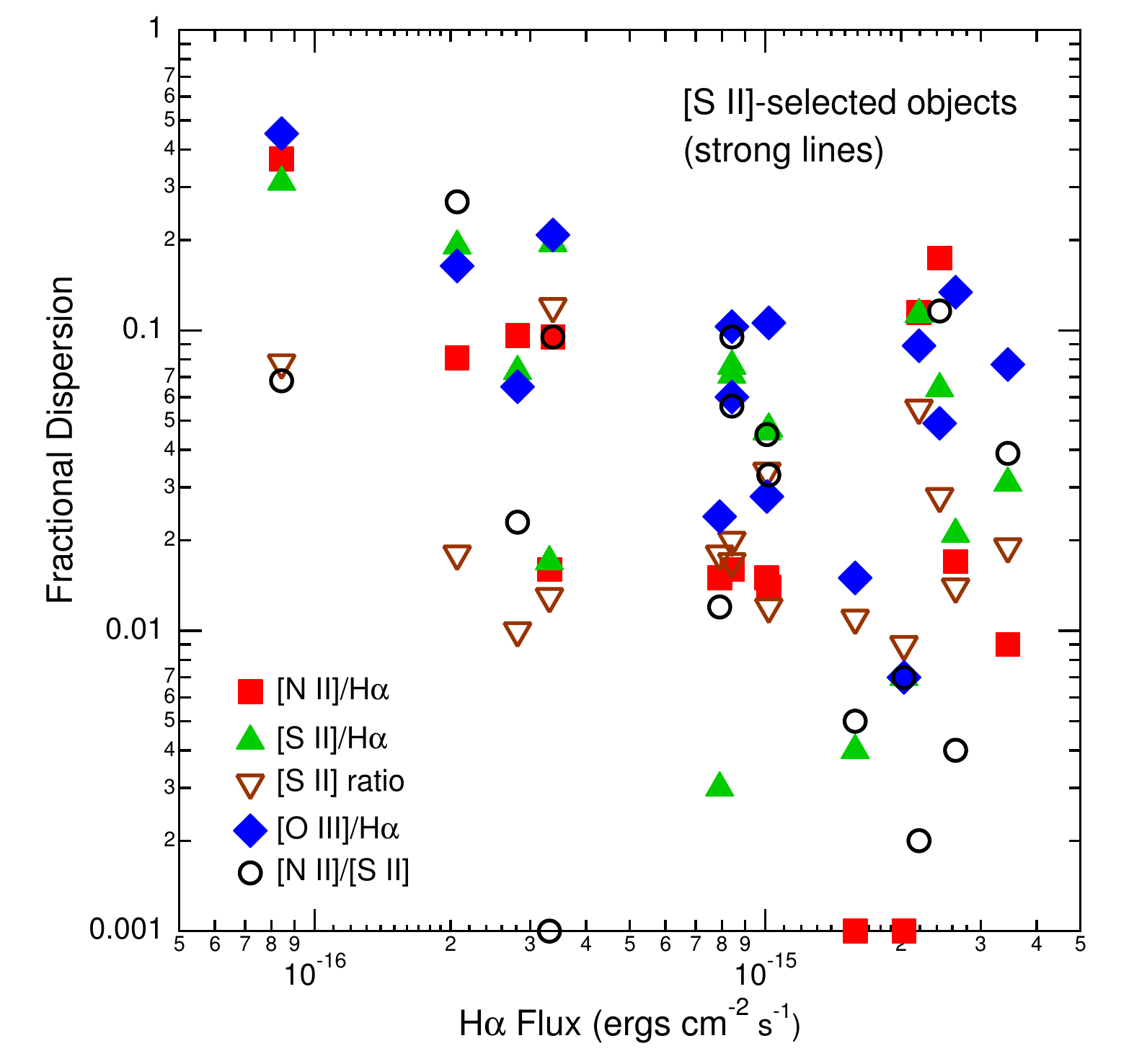}
\caption{Plot of the fractional rms dispersion in values of ratios between various strong lines for the 16 objects with two or more independent GMOS spectra.  For virtually all the objects with \ha\ flux $>2 \times 10^{-16} \FLUX$, the fractional dispersion is$\lesssim 20\%$.}
\label{fig_errors}
\end{figure}

Given this, the best way to estimate the overall uncertainty is to compare completely independent spectra of the same object.  For 16 of our \sii-selected SNRs, we obtained independent spectra from slits on two or more different masks, so we can compare these results.  Even though the integrated flux values in Table \ref{table-snr-flux}  represent only a fraction of the total flux for objects that are considerably larger than the slit width, the {\em relative} fluxes should still be accurate, so we have compared several line flux ratios for independent spectra of the same object by calculating the rms dispersion in each line ratio for all 16 objects with multiple spectra, with the result shown in Fig.~\ref{fig_errors}.  We find that for the strong lines---\oiii \lam 5007, \ha, \nii, and \sii---the dispersion is $< 20\%$ of the ratio value in virtually all cases, and is $<10\%$ in most cases involving remnants with \ha\ flux $\gtrsim 10^{-16} \FLUX$.  Extrapolating from these examples, we estimate that the uncertainty in fluxes for these lines in Tables~\ref{table-snr-flux} and \ref{table-O3-flux} is no worse than than 10-20\%.  For the fainter lines, \oi\ in many cases and \hb\ in some, the uncertainty is probably larger.  Flux values that are likely to have large uncertainties are preceded by an ``$\sim$'' symbol in Tables \ref{table-snr-flux} and \ref{table-O3-flux}.

 
In addition to the SNRs and candidates that are the focus of our study, we also obtained spectra of several \hii\ regions. Table \ref{h2_regions} lists the \hii\ regions and Table \ref{table-HII-flux} shows  extracted fluxes for the \hii\ regions, which are used in some of the comparisons below.

\subsection{Comparison between Images and Spectra}

\begin{figure}
\plotone{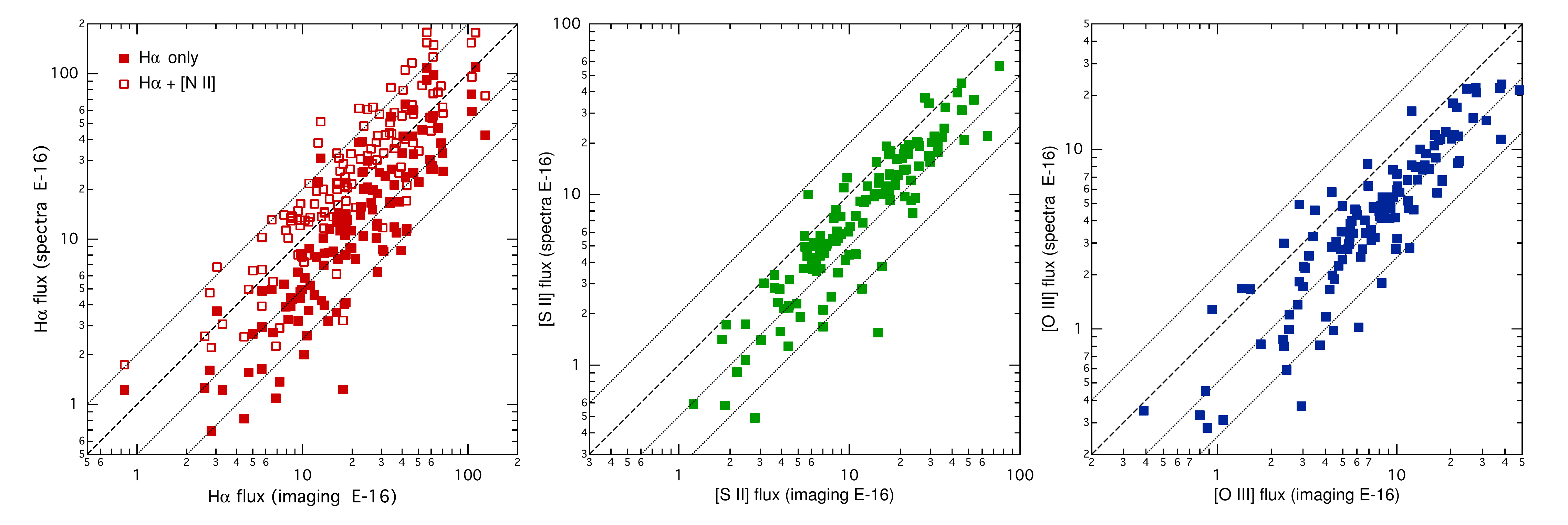}
\caption{({\em left}) Comparison between the \ha\ fluxes measured from narrow-band images (horizontal axis) and those measured from the extracted 1-D spectra (vertical axis).  Also shown (open squares) is the sum of the \ha\ + \nii\ \lamlam 6548, 6583 flux from the spectra, since the narrow-band filter used for imaging passed a fraction of the \nii\ flux.  The heavy dashed line corresponds to equal values, while the lighter ones correspond to $F_{spectra} = 1/4, 1/2,\, {\rm and}\ 2 \times F_{image}$. ({\em center}) Same, but for the \sii\ \lamlam 6716, 6731 flux.    ({\em right})  Same, but for the \oiii\ \lam 5007 flux.  The most extreme point, with $F_{spectra} \ll F_{imaging}$ in all three lines, is a single object that was largely off the GMOS slit.}
\label{fig:flux_comp_trio}
\end{figure}

Before examining the results from our survey, we first check for systematic effects by comparing the emission-line fluxes measured from our spectra with those from our narrow-band imaging with either Magellan/IMACS  (B12), or WFC3 (B14).\footnote{Partially as a result of this comparison, we discovered a significant systematic error in the B12 line fluxes.  We have corrected this in an Erratum, \citet{blair17}, and have used those corrected fluxes for the comparison here.  As noted in that Erratum, the corrected fluxes are also consistent with several independent data sets.}  Fig.~\ref{fig:flux_comp_trio}, shows this comparison for the \ha, \sii, and \oiii\ lines for all the SNR candidates.  For both \sii\ and \oiii, the correlation is good: many objects are close to the line of equal flux, and the majority have flux values from spectroscopy only a little below those from imaging, $0.5\,F_{image} \le F_{spectra} \le F_{image}$.  This is just as expected, since in many cases the extent of the object was greater than the slit width, especially including seeing effects which could cause loss of flux.   For \ha, the correlation is less good because the interference filter used for our Magellan/IMACS imaging (which comprises the vast majority of the objects) was centered  near 6552 \AA, and thus admitted much of the light in the \nii\,\lam6548 line, while the \ha\ line was displaced from the transmission peak (see B12 for further discussion).  Furthermore, for objects in confused regions such as spiral arms or near the nucleus, obtaining an accurate measure of the local sky background was difficult and added uncertainty to the flux measurements from imaging in all three lines. 

\subsection{Confirming Bona Fide SNRs: \sii-Selected Objects}

\begin{figure}
\epsscale{0.8}
\plotone{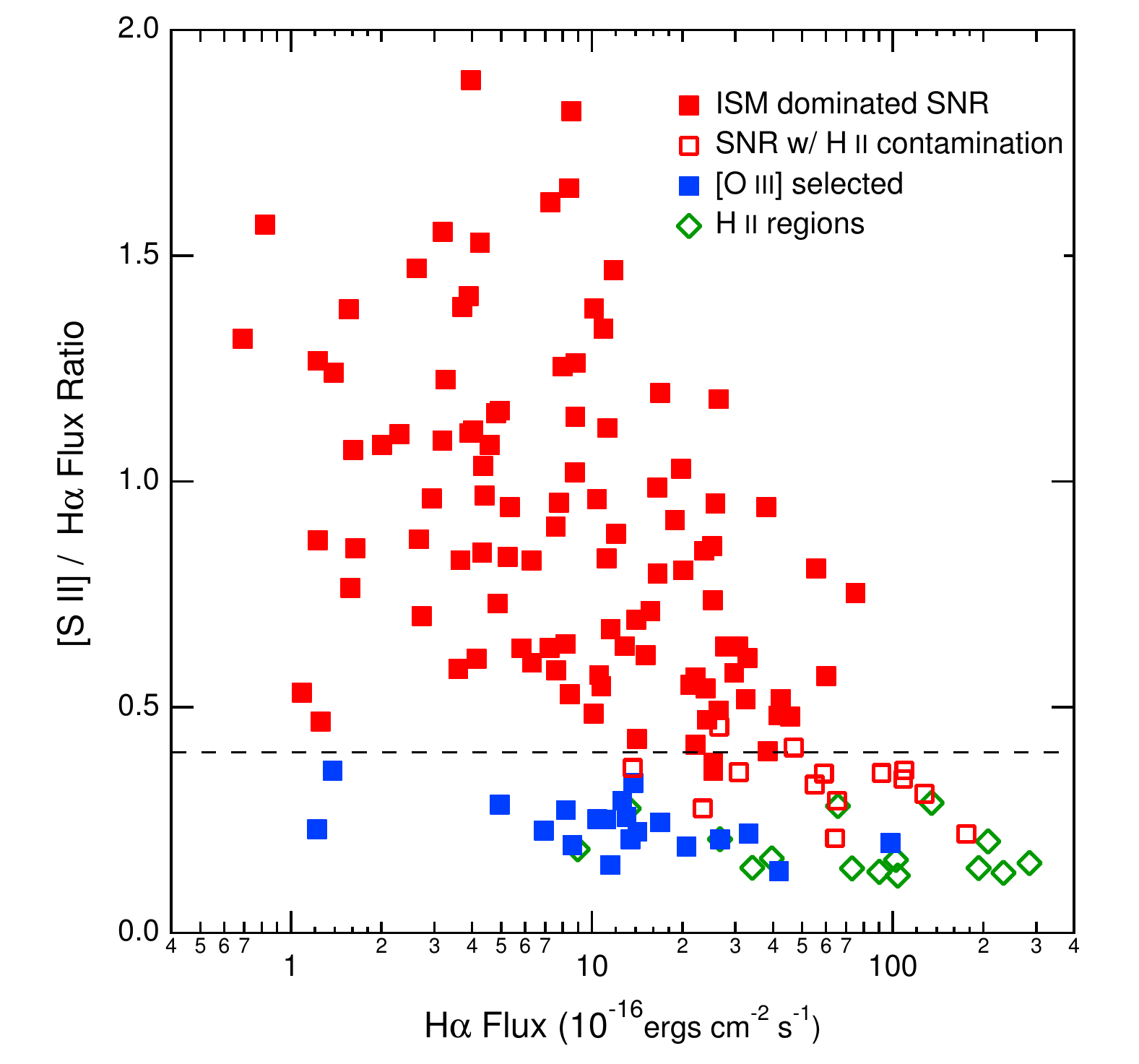}
\caption{Plot of the ratio of \sii\ \lamlam 6716,6731/\ha\ flux as a function of the \ha\ flux level for all the objects we observed spectroscopically.  Filled red squares are SNRs and candidates selected on the basis of high \sii/\ha\ ratios in narrow-band images (Table 2 from B12 and Table 2 from D10).  For virtually all of these, the spectra confirm the high \sii/\ha\ ratio.  Inspection of the WFC3 images for candidates with marginal ratios (\sii/\ha\, $\lesssim 0.4$) shows that most of these are also bona fide SNRs, but with spectra contaminated by coincident \hii\ emission; these are shown as open red squares. Filled blue squares are objects selected on the basis of a high \oiii/\ha\ ratio in images (Table 3 from B12 and  Table 4 from D10).  Also shown are a number of known \hii\ regions targeted in our survey (open green symbols).   Most of the \oiii-selected objects appear to be either PNe or high-excitation \hii\ regions.  No strong trend of the ratio with \ha\ flux is evident. 
}
\label{fig_s2ha_vs_flux}
\end{figure}

One of the goals of this work is to identify which of the SNR candidates we have observed can be confirmed as genuine.  The chief criterion is the \sii/\ha\ ratio, for which only spectra can give an accurate value (since narrow-band images include some or all of the \nii\ \lamlam 6548, 6583 flux along with \ha).  
In Fig.~\ref{fig_s2ha_vs_flux} we plot the \sii/\ha\ ratio as a function of \ha\ flux for various classes of objects in our survey.   For the ISM-dominated candidates---those  that were selected on the basis of high \sii/\ha\ in our narrow-band images---the spectra confirm that this ratio exceeds 0.4, the usual threshold for shock-heated gas, in the vast majority of cases.  The remainder of this section discusses only this group; we return to the \oiii-selected objects in the following section.

An additional  criterion for confirming SNR candidates is the existence of \oi \lamlam 6300, 6363 emission, since neutral oxygen is even less likely to be found in photo-ionized regions than S$^+$.  We find that \oi \lam 6300 is clearly present in the spectra from 110 of the 118 ISM-dominated candidates.
Further characteristics of many, though not all SNRs are the existence of \feii \lam 1.644$\mu$m emission and/or X-ray emission.  The infrared fields of our {\em HST} WFC3 survey of M83 (D10, B14) included 103 of the 118 \sii-selected candidates, of which 56 have obvious coincident 1.644$\mu$m emission in continuum-subtracted F164N images.  We will discuss our WFC3 IR survey of M83 more quantitatively in a future publication.


For only 15 of the ISM-dominated objects  is the \sii/\ha\ ratio measured from spectroscopy smaller than 0.4.  We have re-examined  the Magellan images, those from WFC3 where the object lies in the survey field,  the two-dimensional spectra, and the {\em Chandra} images for all 15 of these, and also for four additional objects where the \sii/\ha\ ratio is only slightly over 0.4.  Almost all of these 19 objects show strong evidence of being SNRs that are contaminated by \hii\ emission---objects either within or adjacent to \hii\ regions---so that the SNR and \hii\ emission could not be separated on the GMOS slit.  All but four of the 19 have clear \oi\lam 6300 emission, and 
nine of them have \feii\lam 1.644 $\mu$m emission detected in our WFC3 IR fields.  Seven of the 19 have coincident soft X-ray sources in our {\em Chandra} survey (L14)\footnote{We only report coincidences between optical SNR candidates and the L14 survey; it is possible that a more thorough approach to measuring the X-ray emission at the positions of SNR candidates might turn up additional coincidences at lower statistical significance.}

For only one object, number 48 from Table 3 of B14 = B14-48, does detailed examination fail to find evidence that the object is a SNR.  This object has a \sii/\ha\ ratio of only 0.06 in the WFC3 images, and 0.13 in its GMOS spectrum. (It was included in the B14 list largely because of its strong \oiii/\ha\ ratio.)  We conclude that it is probably {\em not} a SNR, but instead is a compact \hii\ region, and we exclude it from consideration in the discussion that follows.  For one additional object, B12-119, the evidence that it is a bona fide SNR is somewhat questionable: \sii/H$\alpha \approx 0.37$;  \oi\ emission appears to be present at only  marginal significance; and there is no detected \feii\ or X-ray emission.  Nevertheless, since its  \sii/\ha\ ratio is enhanced relative to the vast majority of \hii\ regions, we retain it as a probable SNR in the subsequent discussion.



\subsection{\oiii-selected Objects}

Also plotted in Fig.~\ref{fig_s2ha_vs_flux} are the 22 objects listed in Table \ref{table-O3} that were selected on the basis of a high ratio of \oiii \lam 5007 to \ha\ flux in narrow-band imaging studies.  The primary purpose of their selection, and of targeting them for spectra, was to search for possible young, ejecta-dominated, or oxygen-rich SNRs like Cas A, where spectra might show broad lines indicating high ejecta velocities.  One such object was indeed found, B12-174a \citep{blair15}.
Also included in the \oiii-selected group is the SN1957D remnant, long recognized as an O-rich SNR.   Of the other 20 objects in this group, none turned out to have the broad emission lines that characterize ejecta-dominated SNRs.   
While the vast majority of H~II regions in M83 have weak or no \oiii\ with respect to H$\beta$, there exists a small subset of compact emission nebulae with stronger \oiii\  emission that are either PNe or compact, high-excitation \hii\ regions (perhaps associated with W-R stars in some cases).  One object selected from the \oiii\  image even turned out to be a background QSO.
Thus, while objects like Cas A or the SN1957D remnant do indeed have high \oiii/\ha\ line ratios, the vast majority of \oiii-bright objects have other identifications, with the stronger than typical \oiii\ emission resulting from photo-ionization.  We conclude that using the \oiii/\ha\ ratio from images to search for ejecta-dominated SNRs is not very efficient, and results in a large number of ``false positives."  We address the question of why M83 has fewer ejecta-dominated remnants than might be expected in Sec.~4.1.


\subsection{Reddening and Emission-Line Ratios} \label{sec:reddening}

\begin{figure}
\epsscale{1.0}
\plotone{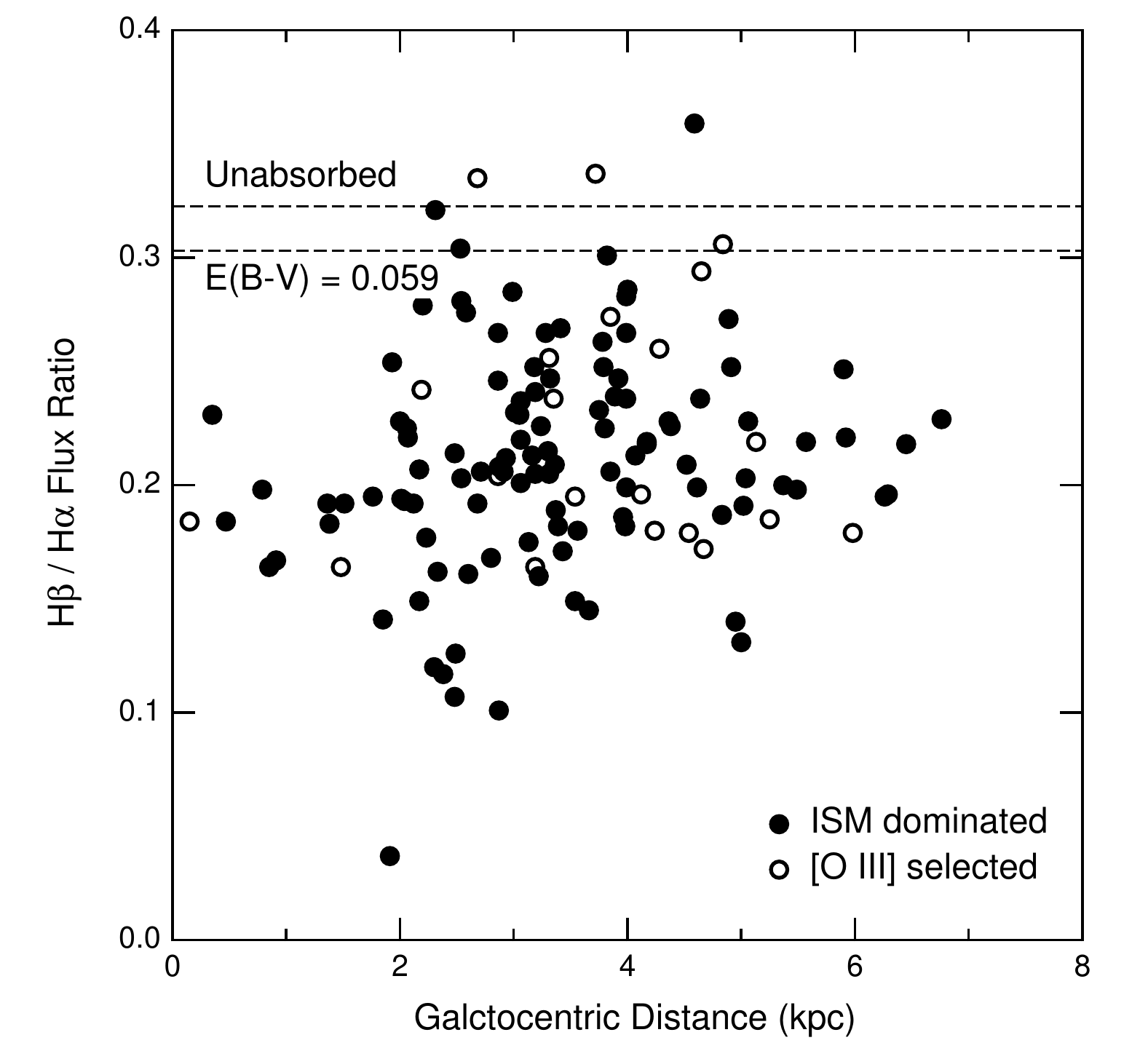}
\caption{Plot of \hb/\ha\ ratio as a function of galactocentric distance for all the SNRs.
The dashed lines indicate the unabsorbed value of \hb/\ha\ = 1/3.1 and the value corresponding to the Galactic foreground absorption, $E(B-V) = 0.059$.  Many of the objects show evidence of signficant local extinction above the foreground level.  However, the ratios we use in our figures are from closely spaced lines with little sensitivity to reddening.}
\label{fig_hbha_vs_r}
\end{figure}

The \hb:\ha\ ratio, plotted in Fig.~\ref{fig_hbha_vs_r}, gives a measure of the reddening, and shows that all our SNR candidates suffer from some degree of local reddening in M83 itself, in addition to the Galactic foreground value of $E(B-V)=0.059$ \citep{schlafly11}, though  the internal reddening generally becomes smaller at larger galactocentric radii.  The range of reddening values is  similar to that found for \hii\ regions in M83 by \citet{bresolin02} and \citet{bresolin05}.  
The reddening is quite uncertain in many cases, due primarily to the difficulty of cleanly subtracting the \ha\ and \hb\ lines in the two-dimensional spectra where there is frequently \hii\ emission along much of the slit.   Hence in Tables \ref{table-snr-flux} and \ref{table-O3-flux} we report the {\em measured}, rather than dereddened fluxes. We note that the most physically important ratios, \sii/\ha, \nii/\ha, \sii \lam 6716:\lam 6731, and \oiii/\hb, the wavelength baselines are quite short, hence these are relatively insensitive to uncertainty in the reddening.  One might suspect that X-ray detected SNRs would tend toward lower values of E(B-V), but we have checked and there is no indication that this is the case.


\section{Emission Line Diagnostics and Comparison with Shock Models}

The relative intensities of various emission lines in both SNRs and  \hii\ regions have long been used, in conjunction with the appropriate models, as diagnostics for the chemical abundances and physical conditions in these nebulae.  A prime example is the \nii/\ha\ ratio, which has often been used as a proxy for the N/H abundance (with appropriate scaling) in both types of nebulae. This is because the N$^+$ ionization potential is close to that of hydrogen, and hence both species populate approximately the same regions \citep[e.g.,][]{blair82}.  In galaxies where both have been observed, a plot showing \nii/\ha\ vs. galactocentric distance (GCD) typically shows parallel tracks for \hii\ regions and SNRs, with SNRs offset to higher values of the ratio. (The \nii/\ha\ ratio is enhanced in SNRs for the same reason that  \sii/\ha\ is enhanced, Sec.~1.) In galaxies with abundance gradients, one usually sees a general decrease in the \nii/\ha\  ratio with increasing GCD, albeit with significant  scatter that may me due to factors such as varying shock conditions \citep{blair97, gordon98, lee15} in additional to observational errors.

\begin{figure}
\epsscale{1.0}
\plotone{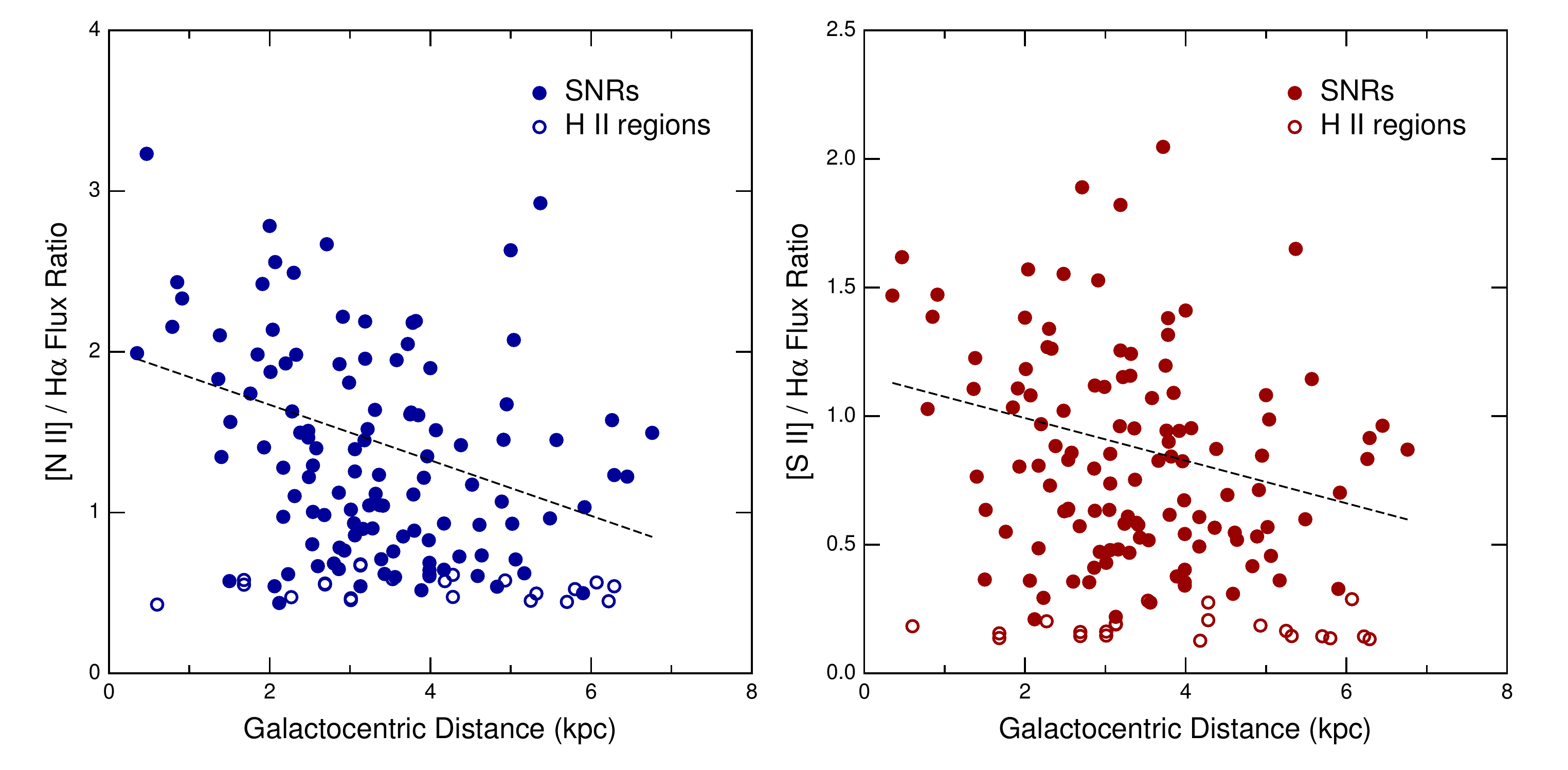}
\caption{Plots of the flux ratio of \nii\,\lamlam 6548,6583/\ha\ (left) and \sii\,\lamlam 6716,6731/\ha\ (right) as a function of deprojected galactocentric distance (GCD) for the ISM-dominated SNRs (filled circles), and also for the \hii\ regions for which we have spectra (open circles). The dashed lines indicate the best linear fits for  ratios for the SNRs, but the implied gradient is of marginal significance given the large scatter in the  ratio values. In contrast, the \hii\ region line ratios are well-behaved and show no evidence of a gradient. The large scatter for the SNRs must result from some combination of varying local abundances and varying shock conditions (see text). 
}
\label{fig_n2_s2_vs_gcd}
\end{figure}

In Fig.~\ref{fig_n2_s2_vs_gcd} we show \nii/\ha\ ratio and \sii/\ha\ ratio for the 
SNRs as a function of GCD.  M83 is unusual among  local spirals in that its mean abundance levels are not only  high, but are {\em uniformly} high over the bright optical disk, with only a  slight metallicity gradient over the bright optical disk  \citep{bresolin02, bresolin09}.
Fig.~\ref{fig_n2_s2_vs_gcd} also includes data for the \hii\ regions (open circles) for which we have spectra (Table \ref{h2_regions}).  The \hii\ regions show little evidence of a radial gradient in either the \nii/\ha\ or \sii/\ha\  ratio, and the scatter is small.  The different between the \hii\ region sample and the SNRs (filled circles) is quite striking. As in other galaxies, the SNRs have significantly higher ratios, but the scatter of the M83 SNR data is quite large.
Fig.~\ref{fig_s2ha_vs_flux} shows that our \hii\ region sample  covers approximately the same range of \ha\ fluxes as the SNR sample, so this is not a signal-to-noise effect. The dashed lines in Fig.~\ref{fig_n2_s2_vs_gcd} show least-squares fits that suggest some decline with GCD in both line ratios for the SNRs, but given the scatter this is only minimally significant. Potential reasons for this large scatter in line ratios will be discussed below, but it likely indicates  both varying shock conditions and real abundance variations among the objects being observed.
 
S$^+$ also has a similar ionization potential to N$^+$ and H, and hence all generally exist together in the same region behind SNR shocks. While the \sii/\ha\ ratio may have some sensitivity to the S/H abundance, it is also more susceptible to varying shock conditions than \nii/\ha.  For instance, the \sii\ doublet ratio, \lam6716/\lam6731, is a well-known indicator of electron density. The density sensitivity arises from the collisional de-excitation of the \lam6716 line relative to \lam6731,  which impacts the ratio of the two lines, but also reduces the total  strength of \sii\ relative to \ha. 
To the extent that the S/H abundance varies among objects, it may arise due to differential grain destruction releasing S from grains (as opposed to N which is not refractory).

Despite these differences, it is an observational fact that the \nii/\ha\ ratio and \sii/\ha\ ratio are correlated at some level in SNRs.  Fig.~\ref{fig_n2_s2_galaxies} shows this correlation for confirmed SNRs in M83 and for those in three other nearby spiral galaxies with good spectrophotometry: M31 \citep{galarza99}, M33 \citep{gordon98}, and M81 \citep{lee15}.  The data for all the galaxies show the correlation, but the mean value is higher in M83 and the range seen in the M83 data is considerably larger than for the other galaxies, as is the scatter in values (especially at the high end of the correlation).  The excess is particularly high for the \nii/\ha\ ratio, reaching more than a factor of two above the comparison galaxies. These trends are generally consistent with the idea that the metallicity of M83 are significantly higher than in the other galaxies, which in turn is consistent with the high abundances inferred for M83 by other methods \citep[][who use several diagnostics to derive O abundance as a proxy for overall metallicity]{bresolin16}.  
Hence, both the large range of observed values of these ratios and the large scatter at a given value of the ratios in M83 need to be understood. 

Since both the \sii/\ha\ and \nii/\ha\ ratios have \ha\ in the denominator, we first examine whether some systematic effect may lead to low \ha\ flux values.  Since there is some \ha\ emission present over much of the galactic disk, it is possible that over- or under-subtracted background emission along the slitlets is responsible for some of the observed scatter.  Since it is difficult to evaluate the effects of background subtraction on the entire set of observations, we have instead inspected two smaller subsets of the data as a  check on the background subtraction.  

The first  was to select a random sample of 15 objects from various slit masks and assess the characteristics and quality of the subtraction, by using {\sc DS9} to inspect the 2-D spectra both before and after background subtraction.   The sample included both bright and faint objects, isolated ones, and and ones with confused background emission. 
 Only a few objects with the most variation in \ha\ background strength along the spatial dimension showed any significant uncertainty in the subtraction at \ha\ at the SNR position.  Thus, while a small fraction of the individual objects may have confusion from structured overlying emission, no overall systematic effect was evident.

The second check was to  inspect the 15 objects with the highest ratios.  If these objects are faint, or ones in highly confused backgrounds, perhaps systematic over-subtraction of background \ha\ might lead to high ratios.  Display of the 2-D spectra showed no apparent systematic effects, and furthermore that several of these objects are well detected  isolated objects with little possible uncertainty in the background subtraction.  The extended high-ratio regime seen for both the \nii/\ha\ ratio and \sii/\ha\ ratio the M83 objects must be a real effect.

\subsection{Insights from Comparison to Shock Models}

Much of what we know from shock-model calculations is due to grids of models at solar or sub-solar abundances, where other parameters like the magnetic field strength or electron densities are held constant \citep[e.g.,][]{raymond79, dopita84, hartigan87, dopita96}.
In this abundance regime, we have come to understand certain properties of the models that can be applied to observations.  For instance, in these models,  the behavior of a line ratio such as \oiii/\hb\ typically increases smoothly as the shock velocity increases from below 100 $\kms$ to 200 $\kms$, that is, as the shocks are able to ionize more and more of the oxygen to and above O$^{++}$. Thus the \oiii/\hb\ line ratio has become a surrogate for shock velocity (at least as to whether it is above or below $\sim 100 \kms$).

The case of M83 is somewhat different situation because of the high metallicity, since few previous shock model grids cover the relevant high abundance regime. The most sophisticated and relevant models in the literature are those of \citet{allen08} produced using the {\sc Mappings III} code.  The model grids provided by these authors cover a significant range of parameter space  that includes a grid of models at twice solar abundance, near that of M83.  However, the shock velocity range covered (100 - 1000 $\kms$) may not be entirely appropriate.  For example, a number of our objects have low \oiii/\hb\ values that probably indicate  shocks well below 100 $\kms$.  On the upper end,  velocities above $\sim$300 $\kms$ are not relevant; otherwise  we would see evidence of line broadening in the spectra. Nonetheless, the trends seen in these model grids  provide some insights into our observations.

There are other factors that limit the applicability of shock models to our data.  Spectra of extragalactic SNRs gather light from most or all of each object, which  averages over fine scale differences that must be present. Also, shock models themselves necessarily involve a number of variables, many of which are mildly if at all constrained by observations (e.g., magnetic field strength, pre-shock ionization conditions, in addition to shock velocity, pre-shock density, and of course, abundances).  Hence, one does not normally expect to match observed line intensities in detail for a given object.  Figure 19 of \citet{allen08} shows how some of the  line ratios  change as some of these parameters are varied, though the models do not extend to the lowest observed values of \oiii/\hb.  

\begin{figure}
\epsscale{0.8}
\plotone{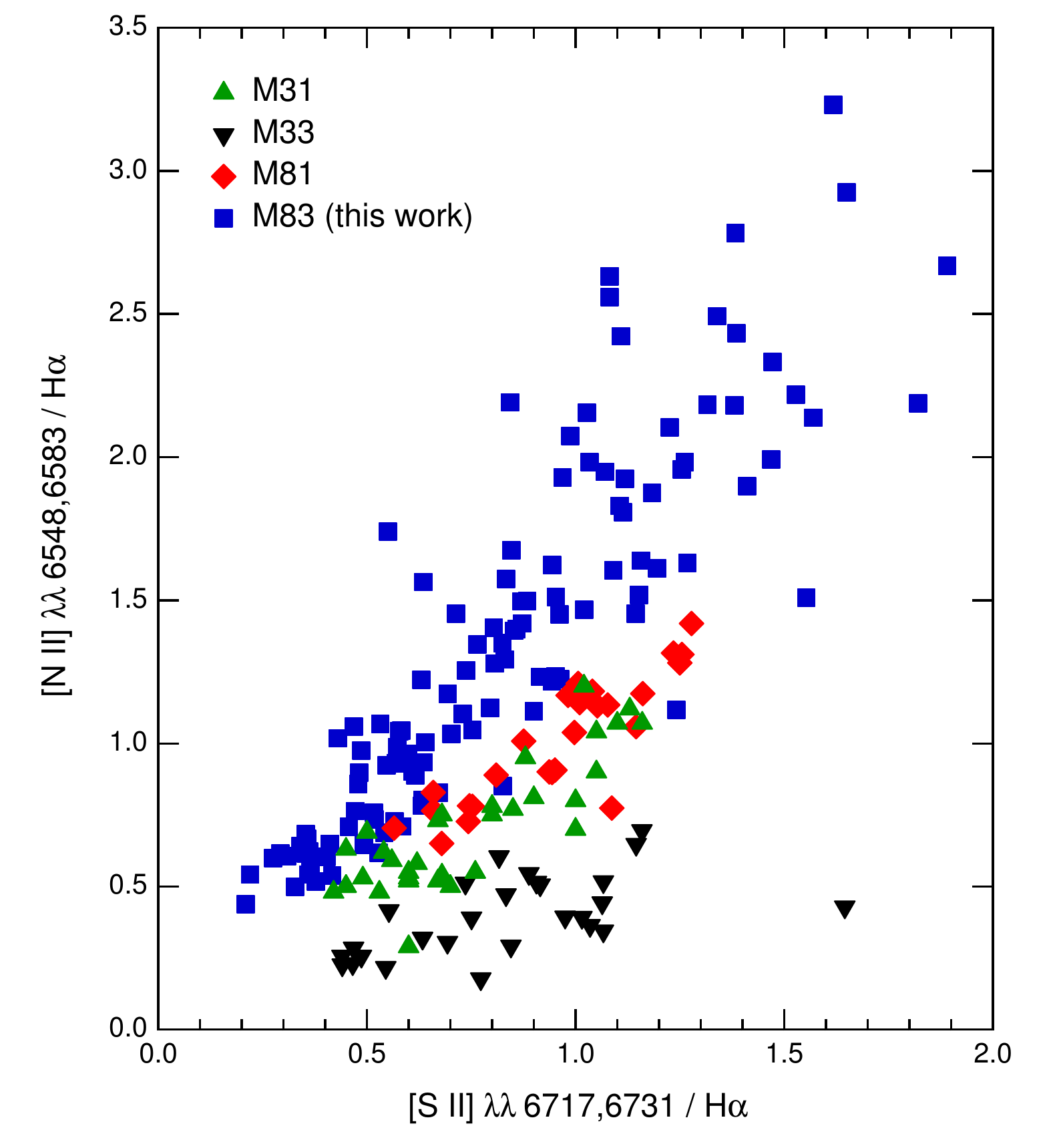}
\caption{This Figure shows the correlation between the \nii\ lines (sum of \lam6548\,+\lam6583 relative to  \ha) and the \sii\ lines (sum of \lam6716\ +\lam6731 relative to  \ha) for confirmed SNRs in M83 and in three other nearby spiral galaxies: M31 \citep{galarza99}, M33 \citep{gordon98}, and M81 \citep{lee15}.  The \nii\ (and in a number of cases \sii) lines are considerably stronger in the M83 objects than for the other galaxies, reflecting the high metallicity of M83.  Furthermore, there is a large dispersion in the \nii/\ha\ line ratios for the M83 objects, which likely indicates the effects of local enhancements in N abundance where SNR shocks encounter circumstellar material that has been enriched due to pre-SN mass loss from the progenitors. }
\label{fig_n2_s2_galaxies}
\end{figure}

\begin{figure}
\begin{centering}
\includegraphics[scale=.37]{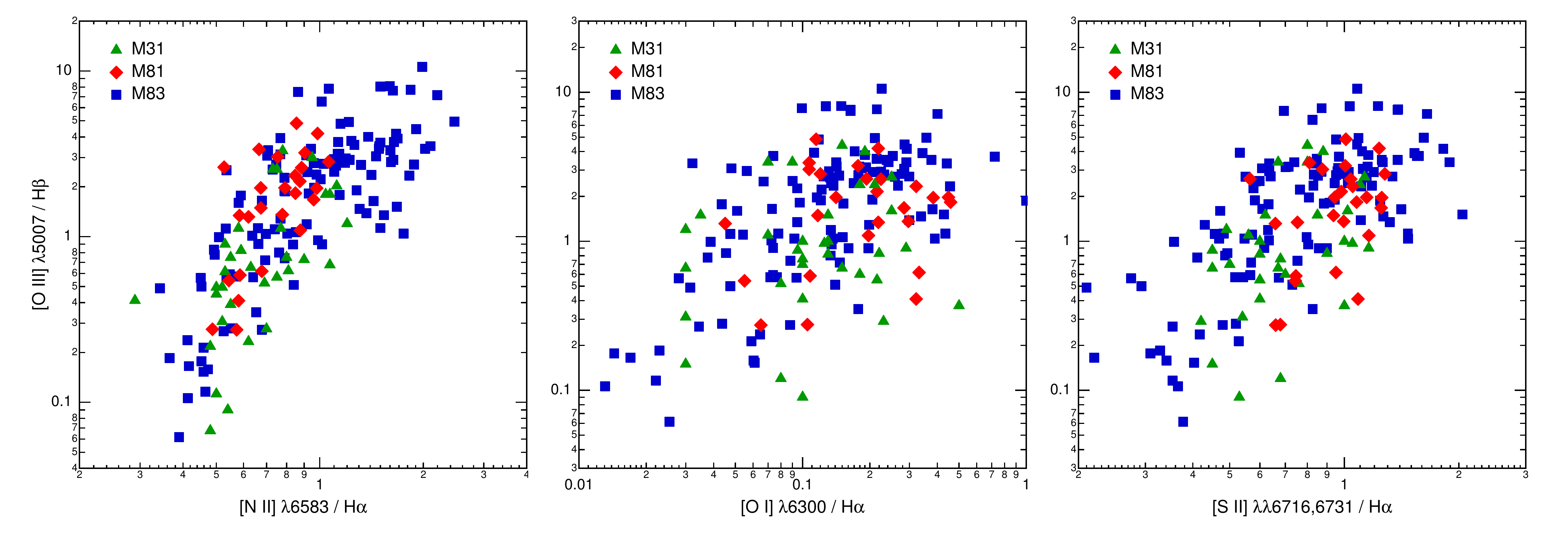}
\caption{Plots showing the flux ratio \oiii\lam5007/\hb, as a function of the following other line ratios for M83 SNRs, and also for ones in M31 \citep{galarza99} and M81 \citep{lee15}: ({\em left})  \nii\lam6583/\ha\ ratio; ({\em center}) \oi\lam6300/\ha\ ratio; ({\em right}) \sii\lamlam6716,6731/\ha\ ratio.  These plots can be compared with corresponding figures in the shock modelling paper of \citet{allen08}.}
\label{fig_o3_hb_galaxies}
\end{centering}
\end{figure}

In Fig.~\ref{fig_o3_hb_galaxies} we show line ratio plots for \oiii/\hb\ as a function of \nii\lam6583/\ha,\footnote{Here we use only the \lam6583 line of \nii, following the usual convention for such diagrams; other plots involving \nii\ in this paper use the sum of the two lines \lamlam6548,6583.} \oi\lam6300/\ha, and \sii\lamlam6716, 6731/\ha---for SNRs in M83, and also for those in M31 and M81---that can be used in comparison with a number of the diagnostic plots shown in \citet{allen08} \citep[see also][]{veilleux87}.  
The most striking feature of these plots is that M83 has many objects with higher \nii/\ha\ ratios than either of the other galaxies, as also seen in Fig.~\ref{fig_n2_s2_galaxies}.
Comparison to \citet{allen08} Fig.~21 shows the effect of increasing abundances, with only the highest-abundance model grids extending to the high \nii/\ha\ values observed in M83.


Some  of observed variation in line ratios is also expected from varying shock conditions;  however, the model sets in Allen et al. cannot explain the high values and large range  in \nii/\ha\ ratio by  varying either the shock velocity or the pre-shock density.
For example, using the 2$\times$ solar model grid from Allen et al.~and a likely range of 100 -- 300 $\kms$ for the models shows a range in \nii/\ha\ ratio from 0.2 -- 0.9, while the observed values extend up to 3 or more in the extreme cases. The \nii/\sii\ ratio shows some variation over this same velocity range, dropping from 1.4 at 100 $\kms$ to 0.9 at 300 $\kms$.  Significantly less variation in this ratio is seen for varying the density (over the same shock velocity range).  Hence, relatively modest changes in the important line ratios are expected due to varying shock conditions.  To summarize then, some of the scatter in \nii/\ha\  for M83 seen in Fig.~\ref{fig_n2_s2_galaxies} may arise from variable shock conditions, but it appears  likely that a high and variable N/H abundance is required to explain the observations.  This is a strong confirmation of the generally high metallicity in M83, with the range in the observed ratio likely reflecting enhanced and variable N abundances {\em locally} in the circumstellar material surrounding many of the individual SNRs.

There are some indications that the abundances in the twice solar models begin to have some unexpected impacts on relative line intensities.  For example, the twice solar models show \oiii/\hb\ as already being strong at 100 $\kms$ and {\em decreasing} dramatically to 200 $\kms$ and above, which is exactly opposite to the lower abundance models over this same range.  Also, more salient to the above discussion, the ratio of \nii/\sii\ varies by over 40\% from 100 -- 300 $\kms$ shocks, whereas this ratio is nearly constant for lower abundance sets over the same velocity range.  As abundances increase, the balance of energy transfer and cooling being carried by the various ions must be changing in ways that are different from the lower-abundance models.  
From even  this cursory comparison, it seems likely  that the elevated abundances may contribute both to the high ratios and to the observed scatter of line ratios we observe in M83.  

Figures 32 and 34 of \citet{allen08}  show their 2$\times$ solar abundance grids projected onto the same line ratio plots as shown in the middle and right panels of Fig.~\ref{fig_o3_hb_galaxies}.  Allen et al.\ apply these models to AGN situations where one attempts to separate shocked emission from photoionized regions, but they should apply to SNR shocks as well.  The majority of the SNRs in all three galaxies plotted in Fig.~\ref{fig_o3_hb_galaxies} lie within the classical shock-heated region toward the upper right in these diagrams, but a significant number of objects appear toward the lower left region nominally ascribed to \hii\ regions.  However, note that the model grids extend significantly down into the nominally photoionized regions of the plots at lower left. (Presumably lower shock velocities might extend even further into this regime.) Here again, we see evidence that at the higher abundances assumed in these models some of the traditional expectations begin to break down. We encourage further investigation of the properties of shock models at elevated abundances that can further elucidate this interesting and so far little-explored region of parameter space.

\subsection{Ramifications}

The high metallicity of M83 has ramifications for the massive stars that are the precursors to many of the SNRs we have observed. It has long been known that abundances affect both the momentum transfer and mass loss in the stellar winds of massive stars \citep[e.g.,][]{vink01,  kudritzki02}, though the situation is complicated \citep{smith14}.
While most such studies consider only solar abundances and below, the trends with increasing abundance are clear, with higher abundances giving rise to significantly stronger and more massive winds.  Coupling this with the hot, high pressure ISM in M83 that was inferred from X-ray data (Long et al. 2014), one might expect this wind mass loss to be constrained to the region immediately surrounding the star, so that when it explodes, the expanding shock would see enhanced density.  This is consistent with the high observed electron densities we report above.

High and variable N abundances in particular might  be expected to arise from such a scenario. As massive stars move beyond H-burning to CNO cycle nucleosynthesis, N becomes enhanced in the atmosphere and hence in the resulting wind \citep[e.g.,][]{{massey00, maeder14}}. Cas A in our Galaxy is one familiar  example of a massive star that exploded with highly enhanced N abundances at its surface (since the fastest moving outer emission knots show high N abundances; Fesen 2001).  Depending on the mass of the precursor star, the amount of mass lost in the wind, and the exact stage at which the SN occurs, one might expect a range of resulting N/H values in the medium surrounding the explosion sites of different stars.  Such a scenario may be at least partially responsible for the wide range of \nii/\ha\ ratios we observe. 

In retrospect, it is interesting to extend this idea to observed SNR populations in other galaxies.  As noted at the beginning of section 4, the observed \nii/\ha\ ratios in other galaxies show significant scatter in addition to possible gradients with GCD 
(see Fig.~9 of Blair \& Long 1997 for NGC~300 and NGC~7793; Fig.~13 of Gordon et al. 1998 for M33; Fig.~21c of Lee et al. 2015 for M31 and M81). While the SNR populations in these other galaxies may not be as dominated by  core-collapse remnants as that in M83,  they still have sufficient star formation and massive star populations that a significant fraction of their SNRs arise from core-collapse events. Thus it seems possible that some of the observed scatter in these galaxies may also arise from local N abundance variations, as opposed to observational scatter and/or variation in shock parameters for the individual objects involved.  Perhaps the more extreme situation we have found in M83 was needed in order to reveal this phenomenon.

\section{The Paucity of Young, Ejecta-Dominated Remnants}
As the host to at least six, and probably seven (counting B12-174a) supernovae within less than a century, we might expect M83 to harbor dozens of SNRs under 1000 years in age.  We would also expect the vast majority of the SNe to have been core-collapse events, like all the  ones with spectra recorded during outburst.   These are  the events that give rise to ejecta-dominated remnants---ones like Cas A \citep[in our Galaxy; age $\sim$335 years,][]{thorstensen01}, E0102--7219 \citep[in the SMC; age $\sim$2000 years,][] {finkelstein06}, or even N132D \citep[in the LMC, which at age $\sim 3000$ years and a diameter of 25 pc still shows evidence of high-velocity O-rich ejecta,][]{morse96}.  All of these objects have spectra that show broad ($\gtrsim 1000 \kms$) emission lines from oxygen and other heavy elements characteristic of SN ejecta.  
Yet we find very few objects with these characteristics in M83: only the remnant of SN1957D \citep[][and references therein]{long12}, and B12-174a, discovered in our survey \citep{blair15}. There can be at most a very few other possible examples, all very small-diameter objects identified in our WFC3 survey (B14) but with no confirming optical spectra to date.  We {\em do} have spectra from over a third of the smallest-diameter (and hence likely youngest) SNRs in our survey, all of which appear to be normal ISM-dominated ones; we address the nature of these in Sec.~5.3.

\subsection{What would known Ejecta-dominated SNRs look like in M83?}

Suppose that analogs to the known ejecta-dominated SNRs {\em were} present in M83; what would they look like, and should we have detected them?  In Table~\ref{table_osnrs} we give the properties of the handful of such objects,  which are also known as oxygen-rich SNRs (OSNRs) because \oiii\lam 5007 is typically the strongest line in their optical spectrum.  There are only three in the Galaxy: Cas A, G292.0+1.8, and Puppis A; three in the Magellanic Clouds: E0102--72.3, N132D, and B0540--69.3; plus the extremely luminous SNR in NGC4449.
For a direct comparison, we also  include in Table~\ref{table_osnrs} SN1957D (in M83), and the Crab Nebula, since it is also the young remnant of a core-collapse SN; though dominated by synchrotron radiation, it also shows emission lines from SN ejecta.

If an M83 analog of Cas A, which at $\sim 335$  years of age has an \oiii\  luminosity of $\sim 2 \times 10^{36}\LUM$, and high ratios of both \oiii/\ha\ and \sii/\ha, it should have been detected in our survey.  Furthermore, it would be very bright in both X-ray and radio bands and have been detected in both of those.   The extraordinary SNR NGC4449-1 has truly extreme luminosities in \oiii, X-rays, and radio; an analog in M83 would have attracted attention even in the earliest resolved surveys.  
The Crab Nebula should also have been detected, both as a radio and an X-ray source.  It is a moderately strong source of optical emission lines (in addition to its synchrotron continuum), so we would have targeted it for a spectrum, and would probably have detected a Crab analog in M83.  Puppis A is bright enough in optical, X-ray and radio bands that it would have been detected, but since the great majority of its optical flux stems from  bright radiative filaments of shocked circumstellar material (CSM), it is unlikely that we would have found the much fainter fast ejecta knots from a Puppis A analog in M83, even in a high signal-to-noise spectrum.   Analogs of E0102--7219 or N132D  might have been detected, but this is not guaranteed.\footnote{While the brightest optical emission from N132D is its outer ``horseshoe" of shocked CSM, the inner ring of shocked ejecta comprises about one-third of the total \oiii\ flux; it should have been detected in WFC3 images (provided, of course, that the source was located within the footprint of the WFC3 fields).}  The \oiii\ flux from both would,   with even the minimal absorption for M83, have placed them in the faintest quartile of our \oiii-selected objects (Table~\ref{table-O3-flux}).  Higher absorption, or location within a confused region, would likely have precluded their detection in our survey.   G292.0+1.8 is bright enough in X-rays that an analog in M83 would probably have been detected in the {\em Chandra} X-ray survey \citep{long14}.   But in \oiii, even the unabsorbed flux from G292 is  fainter than all but one of the objects in Table~\ref{table-O3-flux}.  Despite the fact that it has essentially no \ha\ emission, our optical survey would have detected a G292.0+1.8 analog in M83 only if it were completely isolated, with minimal absorption.  Our survey  would surely not have detected B0540--69.3 were it in M83.

There can be little doubt that Cas A-like objects are extremely under-represented in M83 compared with the number of core-collapse SN events.  More generally, the paucity of Cas A-like SNRs in M33, M31, and other nearby galaxies, now  conclusively including M83, can only mean that Cas A is an exceptional object and not the normal expectation for core-collapse SNRs from massive stars.

\subsection{Why so few?}

Surely many core-collapse SNe have occurred in M83, so we must ask what their remnants must be like, and why so few are ejecta-dominated.  Two possibilities are associated with their environment: (1) For SNe that occur in extremely tenuous surroundings, the resulting blast wave will have little to interact with, will produce only weak reverse shocks propagating into the ejecta, and hence any optical emission will be extremely faint.  
This can happen in situations where multiple SNe have exploded.  The remnant from the first SN would be visible, but subsequent SNRs would be expanding into the evacuated cavity created by the first  explosion.
(2) Perhaps more plausible for M83 is the opposite extreme: that many SNe have exploded in dense environments and have  evolved rapidly  into their radiative phase, in which case fast-moving ejecta knots will be visible for far less than  1000 years.  A high-density environment can occur as  ejecta run into  strong winds from the SN progenitors, or into a dense circumstellar shell---both the result of rapid mass loss at different stages prior to the explosion.   The location of many of the M83 SNRs in young clusters and \hii\ regions likely indicates a generally dense local environment which may also contribute to rapid SNR evolution in these cases.

A closer look  at the known ejecta-dominated SNRs supports the role of surrounding winds and/or shells, and may provide additional insight into why such remnants are so rare.  
 Several, and perhaps all of them, are ones where the ejecta are expanding into a pre-SN stellar wind from the progenitor, or into a cavity carved out by such winds.  The clearest case is for Cas A, where  the light-echo spectrum of the actual SN that shows it to have been a Type IIb event \citep{rest08b, krause08a}.  These are produced from the collapse of the helium core of a red supergiant that had lost most of its hydrogen envelope before exploding, so the ejecta expand into the stellar wind from the pre-SN star \citep{chevalier03, orlando16}.  The extreme luminosity of NGC4449-1 has been best explained by its expansion into a dense and extensive circumstellar environment produced by winds from its massive progenitor, possibly with additional contributions due to winds from other massive stars in the surrounding dense OB cluster \citep{milisavljevic08}.
For G292.0+1.8 and E0102--7219, the fact that fast knots of ejecta are expanding ballistically 2000-3000 years after the explosion requires that they must be expanding into low-density cavities---ones evacuated by pre-SN winds.  In both cases there is an outer shell of X-ray emission where the SN blast wave is interacting with the circumstellar material.  As with Cas A, this interaction also leads to the reverse shock that excites the dense fragments of ejecta, producing the optical emission.
 For both Puppis A and N132D, spectra of the outer radiative filaments show them to be very high in nitrogen; these too are likely overtaking winds enriched by dredged-up nitrogen.  These stripped-envelope SNe---types Ib and Ic---are relatively rare compared to their cousins, types II and IIL, that explode with their envelopes more or less intact; the recent review by \citet{smartt09} indicates that together SN Ib and Ic comprise only $\sim $20-30\% of core-collapse events.   
 
So perhaps the fraction of core-collapse SNe that give rise to ejecta-dominated remnants is relatively small.  And given the dense environments in which most are located, the ones that are produced likely evolve rapidly to the point that the ejecta knots slow, dissipate, and merge  with more normal-abundance material, leading to the large number of ISM-dominated SNRs that are present in M83, even at very small diameters.

\subsection{What are the small-diameter SNRs that we {\em do} see?}

Although we do not see SNRs with the characteristics of Cas A, our total SNR candidate list now contains 41 objects with {\em HST}-measured diameters smaller than $0\farcs 5$ (11 pc), including SN 1957D and B12-174a.
Of the 117 confirmed \sii-selected SNRs with spectra (Table \ref{table-SNR}), 22 are in this small-diameter group.  What are they?

\begin{figure}
\epsscale{0.8}
\plotone{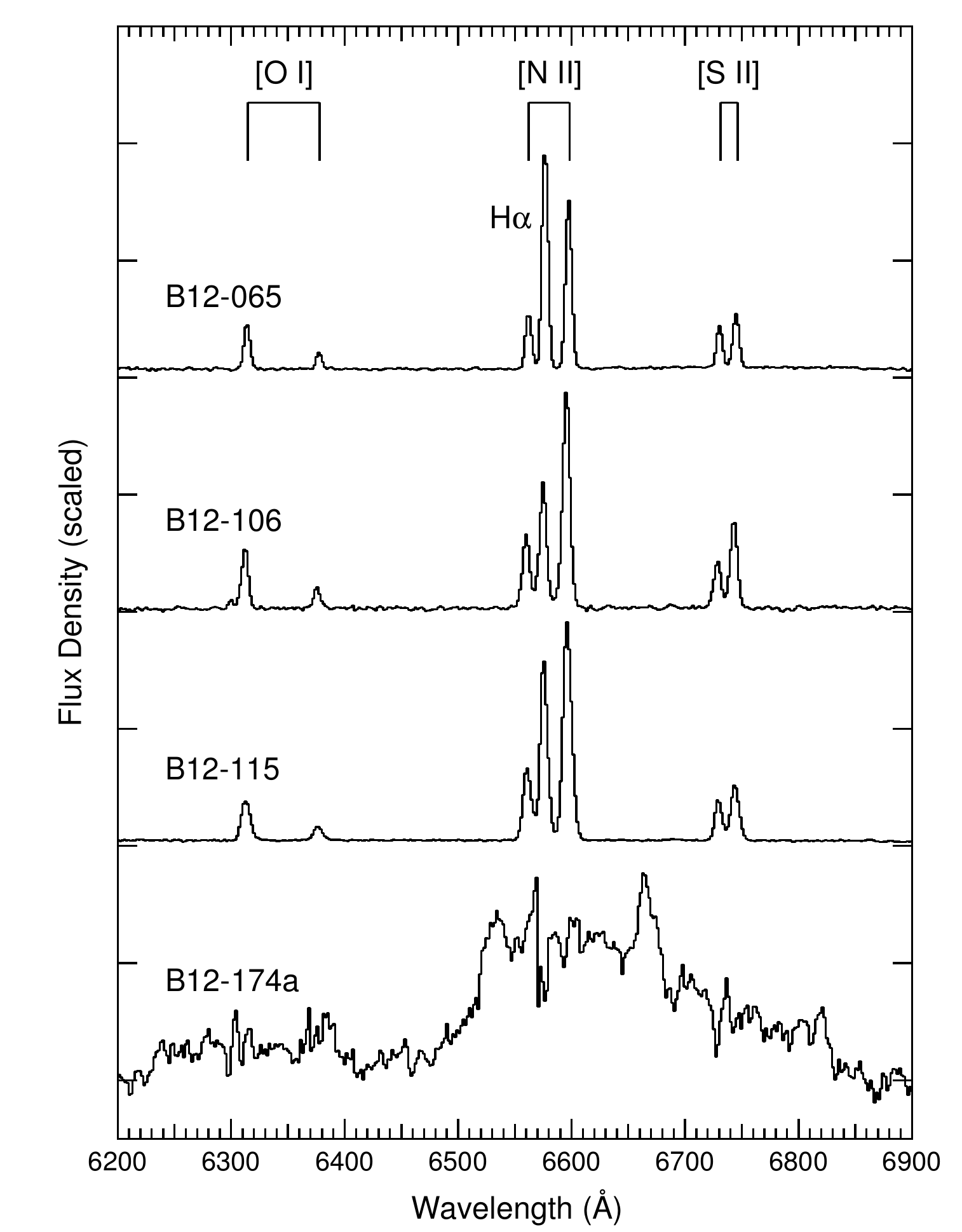}
\caption{GMOS spectra from four of the SNRs with diameter $< 0\farcs 5 = 11$ pc.  B12-174a (bottom trace) is the only one with broad lines characteristic of fast-moving ejecta.  The other three are typical; none of them show  broad lines, and all have \sii  $\lambda 6716/\lambda6731$ ratios indicating relatively high density.  The traces have been scaled and vertically offset for clarity.}
\label{fig_spectra_hi_density}
\end{figure}

\begin{figure}
\epsscale{1.0}

\plotone{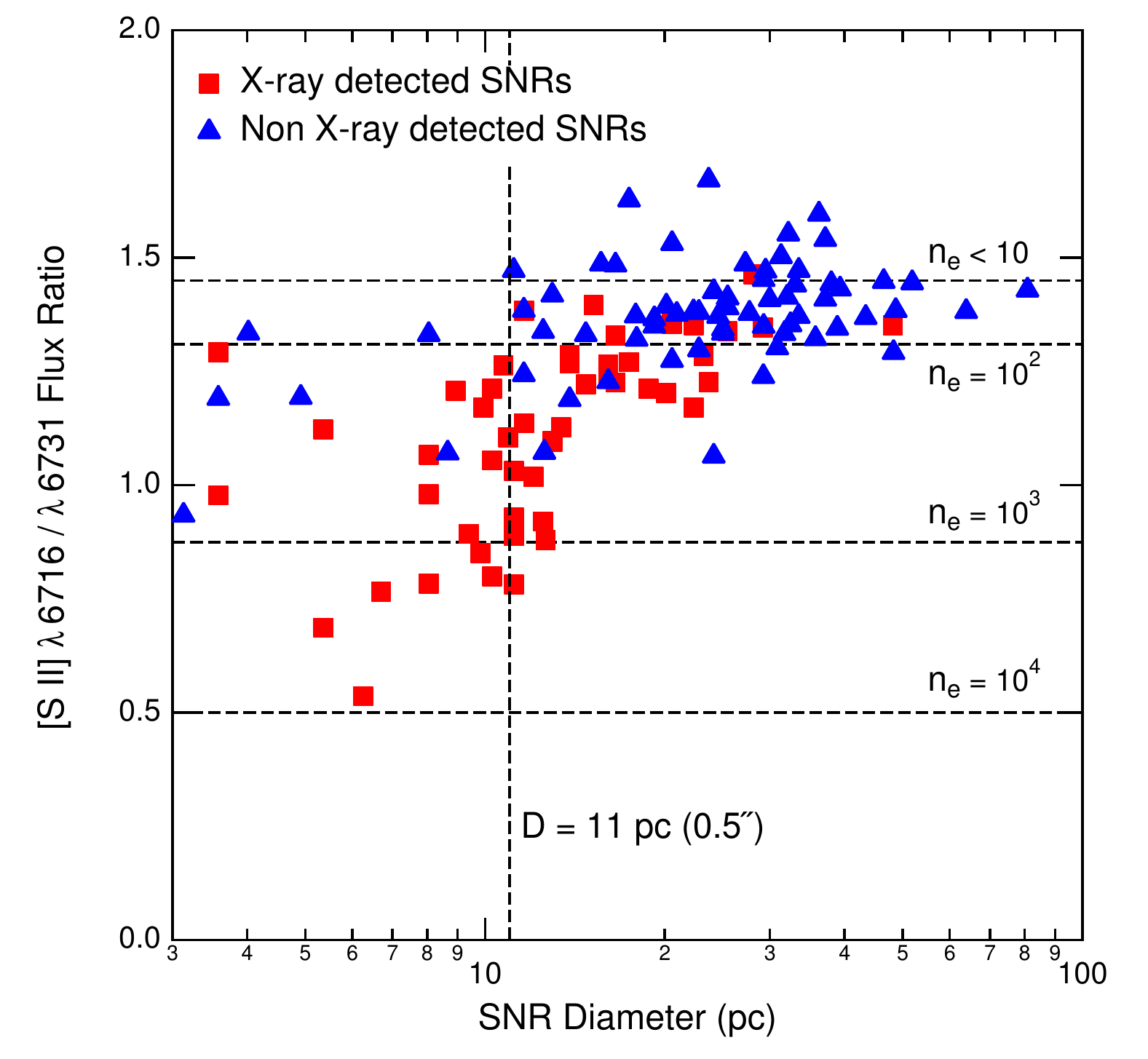}
\caption{Plot of the density-sensitive \sii\ ratio as a function of diameter for all the SNRs, with different symbols for those that have been detected in X-rays (red squares) and those that have not (blue triangles; X-ray data from L14).  Emission from smaller remnants generally stems from higher-density material, suggesting that the smallest (and likely youngest) are evolving into dense circumstellar material, possibly material shed by the precursor stars.  Furthermore, the X-ray detected remnants generally have higher densities, and are smaller (on average) than the others.}
\label{fig_density_vs_diam}
\end{figure}

We show spectra from three of these in Fig.~\ref{fig_spectra_hi_density}; none of these, and  indeed none of the  22, has lines appreciably broader than the instrumental width. (For comparison, the lower trace in Fig.~\ref{fig_spectra_hi_density} shows B12-174a, which {\em does} have broad lines.) But all 22 show fairly high densities, as measured from the \sii\ $\lambda 6716/\lambda6731$ ratio.\footnote{Also in the $D < 11$ pc range are SN1957D and B12-174a, which {\em do} have broad lines and hence high velocities.  We have not included either of these in the Fig.~\ref{fig_density_vs_diam} plot since for both of them the high velocities preclude measurements of the \sii\ ratio.}    

Fig.~\ref{fig_density_vs_diam} shows a plot of the \sii\ density ratio for all of the \sii-selected SNRs plotted as a function of SNR diameter.  The dashed lines provide fiducials for the implied electron densities, with higher densities indicated by smaller values of the ratio. There is a general trend of higher density among smaller diameter SNRs; among the 22 smallest, all have \sii\ ratios measurably different from the low-density limit, and a few approach the high-density limit. Of course, these are the densities in the post-shock zone where S$^+$ is formed, which is compressed by a factor of $\sim$50 from the pre-shock value, depending on variables such as magnetic field strength and other parameters \citep[e.g.,][eq. 8]{dopita79}. Thus, most of these small-diameter objects have inferred pre-shock densities of 10 -- 30 cm$^{-3}$ or more, much higher than typical ISM densities. Since emissivity goes as density squared, these densities refer to the densest pre-shock regions, and this strongly suggests SNRs that are expanding into local density-enhanced regions. The densities in the hot X-ray emitting regions would, of course, be much lower than these values.





What conditions are required to detect these small ISM-dominated SNRs?  In order to be ISM-dominated, the SNR must, by definition, have swept up enough material so that the optical emission from the SNR is not dominated by ejecta emission. Most of the SNRs in M83 will have arisen from core-collapse SNe, and so we expect from 5 to 15 $M_\sun$ of highly-processed material to have been ejected in the SN\@.  In order that emission from a SNR be dominated by optical emission from unprocessed material, the forward shock of the SNR must be propagating either into  dense material from a pre-SN wind or into a very dense interstellar medium.  As a fiducial number, if we assume a SN ejects $10M_\sun$ of material and that the SN shock is propagating into a uniform-density ISM, then to have swept up $10M_\sun$ of material at a diameter of 10 pc, the density of the ISM must be about 0.8\,cm$^{-3}$, certainly not excessive.    
At a diameter of 10 pc, a SNR from an explosion energy of 10$^{51}$ ergs that has reached the Sedov phase would have a primary shock velocity of about 2000 $\VEL$ and an age of 1000 yrs if expanding into an ISM with density 1 cm$^{-3}$; or 650 $\VEL$ and 3000 yrs if the density were 10 cm$^{-3}$.  Essentially no optical emission would arise from this gas, since the cooling time behind the primary shock exceeds the age of the SNR.  

Instead, the optical emission would most likely be arising from slower secondary shocks propagating into dense clumps, and as a result we would expect the velocity broadening of the the optical lines to be much less than the velocity of the primary shock.  For \hii\ regions, the filling factor of dense clumps is of order 1\%, and the densities in the clumps are of order $100\;{\rm cm}^{-3}$ \citep{gutierrez10}.  Once a SNR has reached the Sedov phase, the optical luminosity of a SNR is expected to grow with time, and to be largest once it is in the radiative phase. This favors the detection of SNRs propagating into dense media.  An extreme example is the small (16 pc), bright SNR N49, which is located near a dense cloud in the LMC \citep[e.g.,][]{dopita16}.  Many of the small SNRs in M83 are likely to be objects of this sort.

In retrospect, our hope to identify large numbers of ejecta-dominated SNRs based on a simple extrapolation of the properties of the few of these we know about was probably naive.  Only a small fraction of SNe likely produce such remnants, and the numerous small objects that we do find in M83, while perhaps chronologically young, seem to be ``biologically'' old---dominated by swept-up material from the dense environments in which the explosions occurred.

\section{Summary}


We have obtained spectra of 140 SNR and SNR candidates in M83.  These spectra obtained constitute a representative sample of the 283 emission nebulae identified on the basis of strong \sii/\ha\ ratios from imaging with Magellan and HST, as well as a set of objects suggested to be SNRs on the basis of  strong \oiii\ emission.   Our main conclusions are as follows:

\begin{enumerate}
\item  Nearly all of the emission nebulae identified as SNRs on the basis of large \sii/\ha\ ratios in imaging have similarly high \sii/\ha\ ratios in their spectra.  We have spectroscopically confirmed that 117 objects in M83 are indeed SNRs---the largest uniform sample with spectra in any galaxy.  Only one of the \sii-selected candidates  has been ruled out as a SNR.

\item  None of the 22 \oiii-bright objects (selected in hopes of finding ejected-dominated, O-rich SNRs) for which we have spectra remains as a good SNR candidate, except the small emission nebula identified with the historical SN 1957D \citep{long12} and the recently discovered B12-174a \citep{blair15}. 

\item The \sii/\ha\ and \nii/\ha\ line ratios of the SNRs in M83 are quite high  compared to the values found for SNRs in other spiral galaxies.  The higher values are probably due to the overall metal abundance in M83, and possibly to local enhancements as well.  
The values for these ratios also exceed predictions based on shock models, possibly due to the limited extent to which these models have explored parameter space of gas with super-solar abundances.  Further modeling in this regime would be valuable. 
Both \sii/\ha\ and \nii/\ha\ decrease modestly with galactocentric radius, but with a scatter that is larger than the overall trend.  No other trends of line ratios with galactocentric radius are apparent.  

\item Although there are 68 SNRs in the spectroscopic sample with diameters less than 21 pc, or 1\arcsec, none shows evidence either of broad lines or of interaction with highly enriched ejecta, except the remnant of SN1957D and B12-174a, which, as we have reported in \citet{blair15}, is most likely a the remnant of a SN that exploded within the last 100 years but that was not reported by contemporary observers.  Most of the other small-diameter SNRs are most likely ones that have evolved rapidly following their origin as SNe exploding in relatively dense interstellar environments, similar to N49 in the LMC \citep{dopita16}.

\item A few, but certainly not all, of the handful of known ejecta-dominated SNRs would probably have been detected in our survey had they been located in M83. 
While many dozens of core-collapse SNe must have occurred in M83 over the past millennium, few of these can have produced SNRs like Cas A or E0102--7219.  We attribute this in part to the fact that these known ejecta-dominated remnants resulted from stripped-envelope progenitors, where the ejecta expand into a wind from the pre-SN star. Such progenitors produce SNe of Type Ib/c, which represent only a small fraction of core-collapse SNe.  
In addition, it appears that many of the SNRs in M83 are likely expanding in high-density environments, where remnants  evolve rapidly to the point where they are dominated by swept-up material rather than by ejecta. Others may have exploded in regions where earlier SNRs have evacuated the surrounding region, resulting in very faint SNRs.

\end{enumerate}

\acknowledgements

PFW and WPB acknowledge the excellent  observing support from the staff at the Gemini South Observatory, where the observations reported here were obtained.  
PFW and WPB are grateful for both observing and travel support for the Gemini observations from the Gemini office at NOAO. 
The Gemini Observatory is operated by Association of Universities for Research in Astronomy, Inc., under a cooperative agreement with the NSF on behalf of the Gemini partnership: the National Science Foundation (UnitedStates), the Science and Technology Facilities Council (United Kingdom), the National Research Council (Canada), CONICYT (Chile), the Australian Research Council (Australia),Minist\'{e}rio da Ci\^{e}ncia, Tecnologia e Inova\c{c}\~{a}o (Brazil) and Ministerio de Ciencia, Tecnologia e Innovaci\'{o}n Productiva (Argentina).  In particular, we wish to thank Marcel Bergmann for assistance in refining routines for 1-D spectral extraction.
PFW is grateful for the hospitality of the Research School of Astronomy and Astrophysics, Australian National University, and to valuable discussions with Michael Dopita, during early stages of this work.  PFW also  acknowledges financial support from the National Science Foundation 
through grant AST-0908566 and from NASA through grant HST-GO-12513.03.
WPB acknowledges support from the Dean of the Krieger School of Arts and Sciences and the Center for Astrophysical Sciences at JHU during this work.
Support for this work was also provided by the National Aeronautics and Space Administration 
through \chandra\ Grant Number G01-12115, issued by the \chandra\ X-ray Observatory Center, 
which is operated by the Smithsonian Astrophysical Observatory 
for and on behalf of NASA under contract NAS8-03060\@.  

\facility{Gemini:South (GMOS)},
\facility{Magellan:Baade (IMACS)}, 
\facility{HST (WFC3)} 

\bibliographystyle{apj}

\begin{thebibliography}{82}
\expandafter\ifx\csname natexlab\endcsname\relax\def\natexlab#1{#1}\fi

\bibitem[{{Allen} {et~al.}(2008){Allen}, {Groves}, {Dopita}, {Sutherland}, \&
  {Kewley}}]{allen08}
{Allen}, M.~G., {Groves}, B.~A., {Dopita}, M.~A., {Sutherland}, R.~S., \&
  {Kewley}, L.~J. 2008, \apjs, 178, 20

\bibitem[{{Annibali} {et~al.}(2008){Annibali}, {Aloisi}, {Mack}, {Tosi}, {van
  der Marel}, {Angeretti}, {Leitherer}, \& {Sirianni}}]{annibali08}
{Annibali}, F., {Aloisi}, A., {Mack}, J., {Tosi}, M., {van der Marel}, R.~P.,
  {Angeretti}, L., {Leitherer}, C., \& {Sirianni}, M. 2008, \aj, 135, 1900

\bibitem[{{Barbon} {et~al.}(1999){Barbon}, {Buond{\'{\i}}}, {Cappellaro}, \&
  {Turatto}}]{barbon99}
{Barbon}, R., {Buond{\'{\i}}}, V., {Cappellaro}, E., \& {Turatto}, M. 1999,
  \aaps, 139, 531; current version at heasarc.gsfc.nasa.gov/w3browse/all/asiagosn.html

\bibitem[{{Bigiel} {et~al.}(2010){Bigiel}, {Leroy}, {Seibert}, {Walter},
  {Blitz}, {Thilker}, \& {Madore}}]{bigiel10}
{Bigiel}, F., {Leroy}, A., {Seibert}, M., {Walter}, F., {Blitz}, L., {Thilker},
  D., \& {Madore}, B. 2010, \apjl, 720, L31

\bibitem[{{Blair} {et~al.}(2014){Blair}, {Chandar}, {Dopita}, {Ghavamian},
  {Hammer}, {Kuntz}, {Long}, {Soria}, {Whitmore}, \& {Winkler}}]{blair14}
{Blair}, W.~P., {Chandar}, R., {Dopita}, M.~A., {Ghavamian}, P., {Hammer}, D.,
  {Kuntz}, K.~D., {Long}, K.~S., {Soria}, R., {Whitmore}, B.~C., \& {Winkler},
  P.~F. 2014, \apj, 788, 55

\bibitem[{{Blair} {et~al.}(1982){Blair}, {Kirshner}, \& {Chevalier}}]{blair82}
{Blair}, W.~P., {Kirshner}, R.~P., \& {Chevalier}, R.~A. 1982, \apj, 254, 50

\bibitem[{{Blair} \& {Long}(1997)}]{blair97}
{Blair}, W.~P. \& {Long}, K.~S. 1997, \apjs, 108, 261

\bibitem[{{Blair} \& {Long}(2004)}]{blair04}
---. 2004, \apjs, 155, 101

\bibitem[{{Blair} {et~al.}(2000){Blair}, {Morse}, {Raymond}, {Kirshner},
  {Hughes}, {Dopita}, {Sutherland}, {Long}, \& {Winkler}}]{blair00}
{Blair}, W.~P., {Morse}, J.~A., {Raymond}, J.~C., {Kirshner}, R.~P., {Hughes},
  J.~P., {Dopita}, M.~A., {Sutherland}, R.~S., {Long}, K.~S., \& {Winkler},
  P.~F. 2000, \apj, 537, 667

\bibitem[{{Blair} {et~al.}(1989){Blair}, {Raymond}, {Danziger}, \&
  {Matteucci}}]{blair89}
{Blair}, W.~P., {Raymond}, J.~C., {Danziger}, J., \& {Matteucci}, F. 1989,
  \apj, 338, 812

\bibitem[{Blair {et~al.}(2012)Blair, Winkler, \& Long}]{blair12}
Blair, W.~P., Winkler, P.~F., \& Long, K.~S. 2012, \apjs, 203, 8

\bibitem[{Blair {et~al.}(2017)Blair, Winkler, \& Long}]{blair17}
---. 2017, \apjs, submitted

\bibitem[{{Blair} {et~al.}(2015){Blair}, {Winkler}, {Long}, {Whitmore}, {Kim},
  {Soria}, {Kuntz}, {Plucinsky}, {Dopita}, \& {Stockdale}}]{blair15}
{Blair}, W.~P., {Winkler}, P.~F., {Long}, K.~S., {Whitmore}, B.~C., {Kim}, H.,
  {Soria}, R., {Kuntz}, K.~D., {Plucinsky}, P.~P., {Dopita}, M.~A., \&
  {Stockdale}, C. 2015, \apj, 800, 118

\bibitem[{{Brantseg} {et~al.}(2014){Brantseg}, {McEntaffer}, {Bozzetto},
  {Filipovic}, \& {Grieves}}]{brantseg14}
{Brantseg}, T., {McEntaffer}, R.~L., {Bozzetto}, L.~M., {Filipovic}, M., \&
  {Grieves}, N. 2014, \apj, 780, 50

\bibitem[{{Bresolin} \& {Kennicutt}(2002)}]{bresolin02}
{Bresolin}, F. \& {Kennicutt}, Jr., R.~C. 2002, \apj, 572, 838

\bibitem[{{Bresolin} {et~al.}(2016){Bresolin}, {Kudritzki}, {Urbaneja},
  {Gieren}, {Ho}, \& {Pietrzy{\'n}ski}}]{bresolin16}
{Bresolin}, F., {Kudritzki}, R.-P., {Urbaneja}, M.~A., {Gieren}, W., {Ho},
  I.-T., \& {Pietrzy{\'n}ski}, G. 2016, \apj, 830, 64

\bibitem[{{Bresolin} {et~al.}(2009){Bresolin}, {Ryan-Weber}, {Kennicutt}, \&
  {Goddard}}]{bresolin09}
{Bresolin}, F., {Ryan-Weber}, E., {Kennicutt}, R.~C., \& {Goddard}, Q. 2009,
  \apj, 695, 580

\bibitem[{{Bresolin} {et~al.}(2005){Bresolin}, {Schaerer}, {Gonz{\'a}lez
  Delgado}, \& {Stasi{\'n}ska}}]{bresolin05}
{Bresolin}, F., {Schaerer}, D., {Gonz{\'a}lez Delgado}, R.~M., \&
  {Stasi{\'n}ska}, G. 2005, \aap, 441, 981

\bibitem[{{Chevalier} \& {Oishi}(2003)}]{chevalier03}
{Chevalier}, R.~A. \& {Oishi}, J. 2003, \apjl, 593, L23

\bibitem[{{Dodorico} {et~al.}(1980){Dodorico}, {Dopita}, \&
  {Benvenuti}}]{dodorico80}
{Dodorico}, S., {Dopita}, M.~A., \& {Benvenuti}, P. 1980, \aaps, 40, 67

\bibitem[{{Dopita}(1979)}]{dopita79}
{Dopita}, M.~A. 1979, \apjs, 40, 455

\bibitem[{{Dopita} {et~al.}(2010){Dopita}, {Blair}, {Long}, {Mutchler},
  {Whitmore}, {Kuntz}, {Balick}, {Bond}, {Calzetti}, {Carollo}, {Disney},
  {Frogel}, {O'Connell}, {Hall}, {Holtzman}, {Kimble}, {MacKenty}, {McCarthy},
  {Paresce}, {Saha}, {Silk}, {Sirianni}, {Trauger}, {Walker}, {Windhorst}, \&
  {Young}}]{dopita10}
{Dopita}, M.~A., {Blair}, W.~P., {Long}, K.~S., {Mutchler}, M., {Whitmore},
  B.~C., {Kuntz}, K.~D., {Balick}, B., {Bond}, H.~E., {Calzetti}, D.,
  {Carollo}, M., {Disney}, M., {Frogel}, J.~A., {O'Connell}, R., {Hall}, D.,
  {Holtzman}, J.~A., {Kimble}, R.~A., {MacKenty}, J., {McCarthy}, P.,
  {Paresce}, F., {Saha}, A., {Silk}, J., {Sirianni}, M., {Trauger}, J.,
  {Walker}, A.~R., {Windhorst}, R., \& {Young}, E. 2010, \apj, 710, 964

\bibitem[{{Dopita} {et~al.}(2016){Dopita}, {Seitenzahl}, {Sutherland}, {Vogt},
  {Winkler}, \& {Blair}}]{dopita16}
{Dopita}, M.~A., {Seitenzahl}, I.~R., {Sutherland}, R.~S., {Vogt}, F.~P.~A.,
  {Winkler}, P.~F., \& {Blair}, W.~P. 2016, \apj, 826, 150

\bibitem[{{Dopita} \& {Sutherland}(1996)}]{dopita96}
{Dopita}, M.~A. \& {Sutherland}, R.~S. 1996, \apjs, 102, 161

\bibitem[{{Dopita} \& {Tuohy}(1984)}]{dopita84}
{Dopita}, M.~A. \& {Tuohy}, I.~R. 1984, \apj, 282, 135

\bibitem[{{Elmegreen} {et~al.}(1998){Elmegreen}, {Chromey}, \&
  {Warren}}]{elmegreen98}
{Elmegreen}, D.~M., {Chromey}, F.~R., \& {Warren}, A.~R. 1998, \aj, 116, 2834

\bibitem[{{Eriksen} {et~al.}(2009){Eriksen}, {Arnett}, {McCarthy}, \&
  {Young}}]{eriksen09}
{Eriksen}, K.~A., {Arnett}, D., {McCarthy}, D.~W., \& {Young}, P. 2009, \apj,
  697, 29

\bibitem[{{Fesen}(2001)}]{fesen01}
{Fesen}, R.~A. 2001, \apjs, 133, 161

\bibitem[{{Filipovic} {et~al.}(1998){Filipovic}, {Haynes}, {White}, \&
  {Jones}}]{filipovic98}
{Filipovic}, M.~D., {Haynes}, R.~F., {White}, G.~L., \& {Jones}, P.~A. 1998,
  \aaps, 130, 421

\bibitem[{{Finkelstein} {et~al.}(2006){Finkelstein}, {Morse}, {Green},
  {Linsky}, {Shull}, {Snow}, {Stocke}, {Brownsberger}, {Ebbets}, {Wilkinson},
  {Heap}, {Leitherer}, {Savage}, {Siegmund}, \& {Stern}}]{finkelstein06}
{Finkelstein}, S.~L., {Morse}, J.~A., {Green}, J.~C., {Linsky}, J.~L., {Shull},
  J.~M., {Snow}, T.~P., {Stocke}, J.~T., {Brownsberger}, K.~R., {Ebbets},
  D.~C., {Wilkinson}, E., {Heap}, S.~R., {Leitherer}, C., {Savage}, B.~D.,
  {Siegmund}, O.~H., \& {Stern}, A. 2006, \apj, 641, 919

\bibitem[{{Gaensler} \& {Wallace}(2003)}]{gaensler03}
{Gaensler}, B.~M. \& {Wallace}, B.~J. 2003, \apj, 594, 326

\bibitem[{{Galarza} {et~al.}(1999){Galarza}, {Walterbos}, \&
  {Braun}}]{galarza99}
{Galarza}, V.~C., {Walterbos}, R.~A.~M., \& {Braun}, R. 1999, \aj, 118, 2775

\bibitem[{{Gordon} {et~al.}(1998){Gordon}, {Kirshner}, {Long}, {Blair},
  {Duric}, \& {Smith}}]{gordon98}
{Gordon}, S.~M., {Kirshner}, R.~P., {Long}, K.~S., {Blair}, W.~P., {Duric}, N.,
  \& {Smith}, R.~C. 1998, \apjs, 117, 89

\bibitem[{{Green}(2014)}]{green14}
{Green}, D.~A. 2014, Bulletin of the Astronomical Society of India, 42, 47

\bibitem[{{Guti{\'e}rrez} \& {Beckman}(2010)}]{gutierrez10}
{Guti{\'e}rrez}, L. \& {Beckman}, J.~E. 2010, \apjl, 710, L44

\bibitem[{{Hamuy} {et~al.}(1992){Hamuy}, {Walker}, {Suntzeff}, {Gigoux},
  {Heathcote}, \& {Phillips}}]{hamuy92}
{Hamuy}, M., {Walker}, A.~R., {Suntzeff}, N.~B., {Gigoux}, P., {Heathcote},
  S.~R., \& {Phillips}, M.~M. 1992, \pasp, 104, 533

\bibitem[{{Hartigan} {et~al.}(1987){Hartigan}, {Raymond}, \&
  {Hartmann}}]{hartigan87}
{Hartigan}, P., {Raymond}, J., \& {Hartmann}, L. 1987, \apj, 316, 323

\bibitem[{{Huchtmeier} \& {Bohnenstengel}(1981)}]{huchtmeier81}
{Huchtmeier}, W.~K. \& {Bohnenstengel}, H.-D. 1981, \aap, 100, 72

\bibitem[{{Kirshner} \& {Chevalier}(1977)}]{kirshner77}
{Kirshner}, R.~P. \& {Chevalier}, R.~A. 1977, \apj, 218, 142

\bibitem[{{Krause} {et~al.}(2008){Krause}, {Birkmann}, {Usuda}, {Hattori},
  {Goto}, {Rieke}, \& {Misselt}}]{krause08a}
{Krause}, O., {Birkmann}, S.~M., {Usuda}, T., {Hattori}, T., {Goto}, M.,
  {Rieke}, G.~H., \& {Misselt}, K.~A. 2008, Science, 320, 1195

\bibitem[{{Kudritzki}(2002)}]{kudritzki02}
{Kudritzki}, R.~P. 2002, \apj, 577, 389

\bibitem[{{Lacey} {et~al.}(2007){Lacey}, {Goss}, \& {Mizouni}}]{lacey07}
{Lacey}, C.~K., {Goss}, W.~M., \& {Mizouni}, L.~K. 2007, \aj, 133, 2156

\bibitem[{{Lee} {et~al.}(2010){Lee}, {Park}, {Hughes}, {Slane}, {Gaensler},
  {Ghavamian}, \& {Burrows}}]{lee10}
{Lee}, J., {Park}, S., {Hughes}, J.~P., {Slane}, P.~O., {Gaensler}, B.~M.,
  {Ghavamian}, P., \& {Burrows}, D.~N. 2010, \apj, 711, 861

\bibitem[{{Lee} {et~al.}(2015){Lee}, {Sohn}, {Lee}, {Lim}, {Jang}, {Ko}, {Koo},
  {Hwang}, {Kim}, \& {Park}}]{lee15}
{Lee}, M.~G., {Sohn}, J., {Lee}, J.~H., {Lim}, S., {Jang}, I.~S., {Ko}, Y.,
  {Koo}, B.-C., {Hwang}, N., {Kim}, S.~C., \& {Park}, B.-G. 2015, \apj, 804, 63

\bibitem[{{Levenson} {et~al.}(1995){Levenson}, {Kirshner}, {Blair}, \&
  {Winkler}}]{levenson95}
{Levenson}, N.~A., {Kirshner}, R.~P., {Blair}, W.~P., \& {Winkler}, P.~F. 1995,
  \aj, 110, 739

\bibitem[{{Long} {et~al.}(2012){Long}, {Blair}, {Godfrey}, {Kuntz},
  {Plucinsky}, {Soria}, {Stockdale}, {Whitmore}, \& {Winkler}}]{long12}
{Long}, K.~S., {Blair}, W.~P., {Godfrey}, L.~E.~H., {Kuntz}, K.~D.,
  {Plucinsky}, P.~P., {Soria}, R., {Stockdale}, C.~J., {Whitmore}, B.~C., \&
  {Winkler}, P.~F. 2012, \apj, 756, 18

\bibitem[{{Long} {et~al.}(1989){Long}, {Blair}, \& {Krzeminski}}]{long89}
{Long}, K.~S., {Blair}, W.~P., \& {Krzeminski}, W. 1989, \apjl, 340, L25

\bibitem[{{Long} {et~al.}(2014){Long}, {Kuntz}, {Blair}, {Godfrey},
  {Plucinsky}, {Soria}, {Stockdale}, \& {Winkler}}]{long14}
{Long}, K.~S., {Kuntz}, K.~D., {Blair}, W.~P., {Godfrey}, L., {Plucinsky},
  P.~P., {Soria}, R., {Stockdale}, C., \& {Winkler}, P.~F. 2014, \apjs, 212, 21

\bibitem[{{Maeder} {et~al.}(2014){Maeder}, {Przybilla}, {Nieva}, {Georgy},
  {Meynet}, {Ekstr{\"o}m}, \& {Eggenberger}}]{maeder14}
{Maeder}, A., {Przybilla}, N., {Nieva}, M.-F., {Georgy}, C., {Meynet}, G.,
  {Ekstr{\"o}m}, S., \& {Eggenberger}, P. 2014, \aap, 565, A39

\bibitem[{{Massey} {et~al.}(2000){Massey}, {Waterhouse}, \&
  {DeGioia-Eastwood}}]{massey00}
{Massey}, P., {Waterhouse}, E., \& {DeGioia-Eastwood}, K. 2000, \aj, 119, 2214

\bibitem[{{Mathewson} \& {Clarke}(1972)}]{mathewson72}
{Mathewson}, D.~S. \& {Clarke}, J.~N. 1972, \apjl, 178, L105

\bibitem[{{Matonick} \& {Fesen}(1997)}]{matonick97}
{Matonick}, D.~M. \& {Fesen}, R.~A. 1997, \apjs, 112, 49

\bibitem[{{Mezcua} {et~al.}(2013){Mezcua}, {Lobanov}, \&
  {Mart{\'{\i}}-Vidal}}]{mezcua13}
{Mezcua}, M., {Lobanov}, A.~P., \& {Mart{\'{\i}}-Vidal}, I. 2013, \mnras, 436,
  2454

\bibitem[{{Milisavljevic} \& {Fesen}(2008)}]{milisavljevic08}
{Milisavljevic}, D. \& {Fesen}, R.~A. 2008, \apj, 677, 306

\bibitem[{{Morse} {et~al.}(1996){Morse}, {Blair}, {Dopita}, {Hughes},
  {Kirshner}, {Long}, {Raymond}, {Sutherland}, \& {Winkler}}]{morse96}
{Morse}, J.~A., {Blair}, W.~P., {Dopita}, M.~A., {Hughes}, J.~P., {Kirshner},
  R.~P., {Long}, K.~S., {Raymond}, J.~C., {Sutherland}, R.~S., \& {Winkler},
  P.~F. 1996, \aj, 112, 509

\bibitem[{{Morse} {et~al.}(2006){Morse}, {Smith}, {Blair}, {Kirshner},
  {Winkler}, \& {Hughes}}]{morse06}
{Morse}, J.~A., {Smith}, N., {Blair}, W.~P., {Kirshner}, R.~P., {Winkler},
  P.~F., \& {Hughes}, J.~P. 2006, \apj, 644, 188

\bibitem[{{Morse} {et~al.}(1995){Morse}, {Winkler}, \& {Kirshner}}]{morse95}
{Morse}, J.~A., {Winkler}, P.~F., \& {Kirshner}, R.~P. 1995, \aj, 109, 2104

\bibitem[{{Orlando} {et~al.}(2016){Orlando}, {Miceli}, {Pumo}, \&
  {Bocchino}}]{orlando16}
{Orlando}, S., {Miceli}, M., {Pumo}, M.~L., \& {Bocchino}, F. 2016, \apj, 822,
  22

\bibitem[{{Patnaude} \& {Fesen}(2003)}]{patnaude03}
{Patnaude}, D.~J. \& {Fesen}, R.~A. 2003, \apj, 587, 221

\bibitem[{{Payne} {et~al.}(2004){Payne}, {Filipovi{\'c}}, {Reid}, {Jones},
  {Staveley-Smith}, \& {White}}]{payne04}
{Payne}, J.~L., {Filipovi{\'c}}, M.~D., {Reid}, W., {Jones}, P.~A.,
  {Staveley-Smith}, L., \& {White}, G.~L. 2004, \mnras, 355, 44

\bibitem[{{Petre} {et~al.}(1996){Petre}, {Becker}, \& {Winkler}}]{petre96}
{Petre}, R., {Becker}, C.~M., \& {Winkler}, P.~F. 1996, \apjl, 465, L43+

\bibitem[{{Pilyugin} {et~al.}(2006){Pilyugin}, {Thuan}, \&
  {V{\'{\i}}lchez}}]{pilyugin06a}
{Pilyugin}, L.~S., {Thuan}, T.~X., \& {V{\'{\i}}lchez}, J.~M. 2006, \mnras,
  367, 1139

\bibitem[{{Pilyugin} {et~al.}(2010){Pilyugin}, {V{\'{\i}}lchez}, \&
  {Thuan}}]{pilyugin10}
{Pilyugin}, L.~S., {V{\'{\i}}lchez}, J.~M., \& {Thuan}, T.~X. 2010, \apj, 720,
  1738

\bibitem[{{Raymond}(1979)}]{raymond79}
{Raymond}, J.~C. 1979, \apjs, 39, 1

\bibitem[{{Reed} {et~al.}(1995){Reed}, {Hester}, {Fabian}, \&
  {Winkler}}]{reed95}
{Reed}, J.~E., {Hester}, J.~J., {Fabian}, A.~C., \& {Winkler}, P.~F. 1995,
  \apj, 440, 706

\bibitem[{{Rest} {et~al.}(2008){Rest}, {Welch}, {Suntzeff}, {Oaster},
  {Lanning}, {Olsen}, {Smith}, {Becker}, {Bergmann}, {Challis}, {Clocchiatti},
  {Cook}, {Damke}, {Garg}, {Huber}, {Matheson}, {Minniti}, {Prieto}, \&
  {Wood-Vasey}}]{rest08b}
{Rest}, A., {Welch}, D.~L., {Suntzeff}, N.~B., {Oaster}, L., {Lanning}, H.,
  {Olsen}, K., {Smith}, R.~C., {Becker}, A.~C., {Bergmann}, M., {Challis}, P.,
  {Clocchiatti}, A., {Cook}, K.~H., {Damke}, G., {Garg}, A., {Huber}, M.~E.,
  {Matheson}, T., {Minniti}, D., {Prieto}, J.~L., \& {Wood-Vasey}, W.~M. 2008,
  \apjl, 681, L81

\bibitem[{{Reynoso} {et~al.}(2003){Reynoso}, {Green}, {Johnston}, {Dubner},
  {Giacani}, \& {Goss}}]{reynoso03}
{Reynoso}, E.~M., {Green}, A.~J., {Johnston}, S., {Dubner}, G.~M., {Giacani},
  E.~B., \& {Goss}, W.~M. 2003, \mnras, 345, 671

\bibitem[{{Saha} {et~al.}(2006){Saha}, {Thim}, {Tammann}, {Reindl}, \&
  {Sandage}}]{saha06}
{Saha}, A., {Thim}, F., {Tammann}, G.~A., {Reindl}, B., \& {Sandage}, A. 2006,
  \apjs, 165, 108

\bibitem[{{Schlafly} \& {Finkbeiner}(2011)}]{schlafly11}
{Schlafly}, E.~F. \& {Finkbeiner}, D.~P. 2011, \apj, 737, 103

\bibitem[{{Smartt}(2009)}]{smartt09}
{Smartt}, S.~J. 2009, \araa, 47, 63

\bibitem[{{Smith}(2003)}]{smith03}
{Smith}, N. 2003, \mnras, 346, 885

\bibitem[{{Smith}(2014)}]{smith14}
---. 2014, \araa, 52, 487

\bibitem[{{Soria} \& {Wu}(2003)}]{soria03}
{Soria}, R. \& {Wu}, K. 2003, \aap, 410, 53

\bibitem[{{Thilker} {et~al.}(2005){Thilker}, {Bianchi}, {Boissier}, {Gil de
  Paz}, {Madore}, {Martin}, {Meurer}, {Neff}, {Rich}, {Schiminovich},
  {Seibert}, {Wyder}, {Barlow}, {Byun}, {Donas}, {Forster}, {Friedman},
  {Heckman}, {Jelinsky}, {Lee}, {Malina}, {Milliard}, {Morrissey}, {Siegmund},
  {Small}, {Szalay}, \& {Welsh}}]{thilker05}
{Thilker}, D.~A., {Bianchi}, L., {Boissier}, S., {Gil de Paz}, A., {Madore},
  B.~F., {Martin}, D.~C., {Meurer}, G.~R., {Neff}, S.~G., {Rich}, R.~M.,
  {Schiminovich}, D., {Seibert}, M., {Wyder}, T.~K., {Barlow}, T.~A., {Byun},
  Y., {Donas}, J., {Forster}, K., {Friedman}, P.~G., {Heckman}, T.~M.,
  {Jelinsky}, P.~N., {Lee}, Y., {Malina}, R.~F., {Milliard}, B., {Morrissey},
  P., {Siegmund}, O.~H.~W., {Small}, T., {Szalay}, A.~S., \& {Welsh}, B.~Y.
  2005, \apjl, 619, L79

\bibitem[{{Thorstensen} {et~al.}(2001){Thorstensen}, {Fesen}, \& {van den
  Bergh}}]{thorstensen01}
{Thorstensen}, J.~R., {Fesen}, R.~A., \& {van den Bergh}, S. 2001, \aj, 122,
  297

\bibitem[{{Trimble}(1973)}]{trimble73}
{Trimble}, V. 1973, \pasp, 85, 579

\bibitem[{{van den Bergh}(1989)}]{vandenbergh89}
{van den Bergh}, S. 1989, \aapr, 1, 111

\bibitem[{{Veilleux} \& {Osterbrock}(1987)}]{veilleux87}
{Veilleux}, S. \& {Osterbrock}, D.~E. 1987, \apjs, 63, 295

\bibitem[{{Vink} {et~al.}(2001){Vink}, {de Koter}, \& {Lamers}}]{vink01}
{Vink}, J.~S., {de Koter}, A., \& {Lamers}, H.~J.~G.~L.~M. 2001, \aap, 369, 574

\bibitem[{{Williams} {et~al.}(2015){Williams}, {Bonanos}, {Whitmore}, {Prieto},
  \& {Blair}}]{williams15}
{Williams}, S.~J., {Bonanos}, A.~Z., {Whitmore}, B.~C., {Prieto}, J.~L., \&
  {Blair}, W.~P. 2015, \aap, 578, A100

\bibitem[{{Winkler} \& {Long}(2006)}]{winkler06}
{Winkler}, P.~F. \& {Long}, K.~S. 2006, \aj, 132, 360

\bibitem[{{Winkler} {et~al.}(1988){Winkler}, {Tuttle}, {Kirshner}, \&
  {Irwin}}]{winkler88}
{Winkler}, P.~F., {Tuttle}, J.~H., {Kirshner}, R.~P., \& {Irwin}, M.~J. 1988,
  in IAU Colloq. 101: Supernova Remnants and the Interstellar Medium, ed. R.~S.
  {Roger} \& T.~L. {Landecker}, 65--+

\bibitem[{{Winkler} {et~al.}(2009){Winkler}, {Twelker}, {Reith}, \&
  {Long}}]{winkler09}
{Winkler}, P.~F., {Twelker}, K., {Reith}, C.~N., \& {Long}, K.~S. 2009, \apj,
  692, 1489

\end{thebibliography}

\clearpage




\begin{deluxetable}{cccl}
\tablewidth{0pt}
\tablecaption{Gemini-S/GMOS Multi-Object Spectroscopy Observations of M83\tablenotemark{a}}

\tablehead{
\colhead{Mask No.} &
\colhead{Date (UT)}  &
\colhead{Exposure (s)} &
\colhead{Comments} 
}

\startdata
 &  & $7 \times 1000$ &    \\[-9pt]
 & 7 Apr 2011 &  &  Variable clouds and seeing  \\[-9pt]
2 &    &  $2 \times 1500$ &  \\[-1pt]
 &  9 Apr 2011  & $2 \times 1400$ & Clear; seeing $<1\arcsec$  \\[2pt]
 1 & 8 Apr 2011 & $3 \times 2400$ &  Clear; seeing $<1\arcsec$ \\[2pt]
4 & 8 Apr 2011 & $3 \times 2000$ &  Clear; seeing $<1\arcsec$  \\[2pt]
  & 8 Apr 2011 & $2 \times 1800$ &  Clear; seeing $<1\arcsec$  \\[-9pt]
5 &  &  &    \\[-9pt]
 & 9 Apr 2011 & $1 \times 2400$ &  Clear; seeing $<1\arcsec$  \\[2pt]
3 & 9 Apr 2011 & $3 \times 2000$ &  Clear; seeing $\lesssim 0\farcs 8$  \\[2pt]
6 & 9 Apr 2011 & $3 \times 2000$ &  Clear; seeing $\lesssim 0\farcs 6$  \\[2pt]
7 & 9 Apr 2011 & $3 \times 1900$ &  Clear; seeing $\lesssim 0\farcs 6$  \\[4pt]
\hline\\[-11pt]
8 & 26 Apr\,-\,19 May 2015 & $15 \times 1000$ & Queue program \\[2pt]
9 & 12, 20 Jun 2015 & $15 \times 1000$ & Queue program \\
\enddata

\tablenotetext{a}{2011 observations were a classical program with e2v chips; 2015 ones were queue-scheduled, with Hamamatsu chips.}

\label{obs_log}
\end{deluxetable}

\begin{deluxetable}{rlccccccl}
\tablecaption{[S II]-selected SNR candidates with Spectra }
\tablehead{\colhead{Name$^a$} & 
 \colhead{Other~Names} & 
 \colhead{R.A.\,(J2000.)} & 
 \colhead{Decl.\,(J2000.)} & 
 \colhead{Diam.} & 
 \colhead{GC~distance} & 
 \colhead{X-ray~detection$^b$} & 
 \colhead{Mask~ID} 
\\
\colhead{~} & 
 \colhead{~} & 
 \colhead{(h:m:s)} & 
 \colhead{(d:m:s)} & 
 \colhead{(pc)} & 
 \colhead{(kpc)} & 
 \colhead{~} & 
 \colhead{~} 
}
\tabletypesize{\scriptsize}
\tablewidth{0pt}\startdata
B12-001 &  - &  13:36:39.99 &  -29:51:35.1 &  33.5 &  6.5 &  - &  2-02 \\ 
B12-003 &  - &  13:36:40.90 &  -29:51:17.7 &  12.5 &  6.3 &  L14-019 &  1-21 \\ 
B12-005 &  - &  13:36:41.58 &  -29:49:56.3 &  13.9 &  6.8 &  - &  1-23 \\ 
B12-010 &  - &  13:36:44.64 &  -29:50:33.9 &  38.0 &  5.5 &  - &  1-19 \\ 
B12-012 &  - &  13:36:45.66 &  -29:52:21.3 &  80.9 &  4.6 &  - &  2-06 \\ 
B12-014\tablenotemark{c} &  - &  13:36:46.42 &  -29:53:42.3 &  38.9 &  4.9 &  - &  2-07 \\ 
B12-020 &  - &  13:36:47.83 &  -29:51:18.1 &  33.1 &  4.2 &  - &  1-17 \\ 
B12-021 &  - &  13:36:47.93 &  -29:51:46.0 &  48.7 &  4.0 &  - &  2-08 \\ 
B12-022 &  - &  13:36:48.11 &  -29:51:33.9 &  33.5 &  4.0 &  - &  8-01 \\ 
B12-023 &  - &  13:36:48.30 &  -29:52:44.7 &  20.1 &  3.9 &  L14-046 &  6-21 \\ 
B12-025 &  - &  13:36:48.59 &  -29:52:05.3 &  51.9 &  3.8 &  - &  4-25 \\ 
B12-026 &  B14-01 &  13:36:48.99 &  -29:52:54.1 &  4.9 &  3.8 &  - &  8-02 \\ 
B12-028 &  - &  13:36:49.37 &  -29:53:19.9 &  39.3 &  3.9 &  - &  6-22 \\ 
B12-031 &  - &  13:36:49.62 &  -29:53:13.7 &  36.2 &  3.8 &  - &  8-03 \\ 
B12-033 &  - &  13:36:49.68 &  -29:54:04.2 &  22.8 &  4.4 &  - &  4-27 \\ 
B12-034 &  - &  13:36:49.72 &  -29:50:57.1 &  46.5 &  3.8 &  - &  1-16 \\ 
B12-036 &  B14-02 &  13:36:49.81 &  -29:52:16.9 &  3.6 &  3.4 &  L14-053 &  2-12 \\ 
B12-035 &  - &  13:36:49.81 &  -29:53:08.3 &  31.3 &  3.7 &  - &  6-24b \\ 
B12-037 &  B14-04 &  13:36:50.12 &  -29:53:08.7 &  8.1 &  3.6 &  L14-057 &  6-24a \\ 
B12-039 &  - &  13:36:50.28 &  -29:52:47.5 &  22.8 &  3.4 &  - &  4-23 \\ 
B12-043 &  - &  13:36:50.76 &  -29:53:10.6 &  27.3 &  3.4 &  - &  6-23 \\ 
B12-045 &  - &  13:36:50.85 &  -29:52:39.6 &  10.3 &  3.2 &  L14-063 &  8-04 \\ 
B12-048 &  - &  13:36:50.99 &  -29:52:25.9 &  48.2 &  3.0 &  L14-065 &  4-24 \\ 
B14-07 &  - &  13:36:51.19 &  -29:50:42.3 &  5.4 &  3.6 &  L14-067 &  1-13 \\ 
B14-09 &  - &  13:36:51.53 &  -29:53:00.9 &  16.1 &  3.1 &  - &  8-21 \\ 
B12-058 &  - &  13:36:52.38 &  -29:52:05.2 &  11.2 &  2.6 &  - &  2-10 \\ 
B12-057 &  - &  13:36:52.40 &  -29:50:43.9 &  31.7 &  3.2 &  - &  1-14 \\ 
B12-065 &  B14-11 &  13:36:53.23 &  -29:53:25.3 &  6.7 &  3.0 &  L14-105 &  4-21 \\ 
B12-067 &  B14-12 &  13:36:53.29 &  -29:52:48.2 &  10.7 &  2.5 &  L14-106 &  6-19 \\ 
B12-066 &  - &  13:36:53.30 &  -29:52:42.5 &  11.2 &  2.5 &  L14-107 &  2-15 \\ 
B12-069 &  - &  13:36:53.37 &  -29:50:38.4 &  12.5 &  3.1 &  - &  1-15 \\ 
B12-073 &  - &  13:36:53.89 &  -29:48:48.2 &  11.6 &  5.1 &  L14-116 &  1-09 \\ 
B12-074 &  - &  13:36:54.16 &  -29:52:09.3 &  16.1 &  2.1 &  L14-119 &  2-11 \\ 
B12-077 &  - &  13:36:54.45 &  -29:56:00.4 &  16.5 &  5.9 &  - &  5-03 \\ 
B12-084 &  - &  13:36:54.87 &  -29:50:18.6 &  23.7 &  3.1 &  L14-128 &  1-08 \\ 
B12-087 &  - &  13:36:55.03 &  -29:52:39.6 &  13.0 &  2.0 &  L14-129 &  2-14 \\ 
B12-089 &  - &  13:36:55.06 &  -29:53:04.5 &  9.8 &  2.3 &  L14-131 &  6-13a \\ 
B12-091 &  - &  13:36:55.22 &  -29:53:05.0 &  23.7 &  2.3 &  - &  6-13b \\ 
B12-097 &  - &  13:36:55.48 &  -29:52:43.7 &  11.6 &  1.9 &  - &  8-05 \\ 
B12-098 &  - &  13:36:55.62 &  -29:53:03.6 &  17.9 &  2.2 &  - &  7-01 \\ 
B12-101 &  - &  13:36:55.80 &  -29:51:19.7 &  37.1 &  1.9 &  - &  1-10 \\ 
B12-104 &  - &  13:36:56.06 &  -29:56:05.7 &  37.1 &  5.9 &  - &  5-05 \\ 
B12-106 &  B14-16 &  13:36:56.23 &  -29:52:55.2 &  6.3 &  1.9 &  L14-141 &  6-16 \\ 
B12-109 &  B14-17 &  13:36:56.81 &  -29:49:49.7 &  11.2 &  3.3 &  L14-149 &  1-11 \\ 
B12-110 &  - &  13:36:56.82 &  -29:49:24.8 &  63.9 &  3.9 &  - &  3-22 \\ 
B12-112 &  - &  13:36:57.16 &  -29:53:33.7 &  32.0 &  2.5 &  - &  4-19 \\ 
B12-115 &  B14-18 &  13:36:57.88 &  -29:53:02.8 &  5.4 &  1.8 &  L14-159 &  4-17 \\ 
B12-117 &  - &  13:36:58.55 &  -29:48:19.7 &  20.5 &  5.2 &  L14-166 &  9-03 \\ 
B14-19 &  - &  13:36:58.64 &  -29:51:06.5 &  4.0 &  1.4 &  - &  8-22 \\ 
B12-118 &  D10-01 &  13:36:58.72 &  -29:51:00.6 &  16.5 &  1.5 &  L14-172 &  3-21 \\ 
B14-20 &  D10-02 &  13:36:58.90 &  -29:52:26.3 &  13.0 &  0.9 &  - &  6-15 \\ 
B12-119 &  B14-21 &  13:36:59.00 &  -29:52:56.8 &  3.6 &  1.5 &  - &  9-04 \\ 
B12-122 &  - &  13:36:59.33 &  -29:55:08.9 &  9.9 &  4.5 &  L14-183 &  5-08 \\ 
B12-125 &  - &  13:36:59.47 &  -29:49:16.6 &  29.3 &  3.8 &  - &  1-07 \\ 
B12-124 &  D10-04 &  13:36:59.50 &  -29:52:03.9 &  12.1 &  0.5 &  L14-186 &  4-10 \\ 
B12-127 &  - &  13:36:59.85 &  -29:55:26.0 &  10.3 &  4.9 &  L14-195 &  7-03 \\ 
B12-129 &  B14-26 &  13:37:00.03 &  -29:54:16.9 &  3.6 &  3.3 &  L14-199 &  4-12 \\ 
B12-130 &  - &  13:37:00.09 &  -29:48:40.3 &  15.6 &  4.6 &  - &  9-05 \\ 
B12-132 &  D10-07 &  13:37:00.34 &  -29:51:20.7 &  20.1 &  0.8 &  - &  1-05 \\ 
B12-133 &  - &  13:37:00.40 &  -29:53:22.9 &  18.8 &  2.0 &  L14-215 &  7-07a \\ 
B12-134 &  - &  13:37:00.68 &  -29:54:26.7 &  25.5 &  3.5 &  - &  4-11 \\ 
B12-136 &  - &  13:37:00.73 &  -29:53:23.7 &  29.9 &  2.1 &  - &  7-07b \\ 
B12-137 &  B14-35;~D10-09 &  13:37:01.02 &  -29:50:56.3 &  8.1 &  1.4 &  L14-235 &  1-04 \\ 
B12-142 &  - &  13:37:01.66 &  -29:54:10.1 &  13.9 &  3.2 &  L14-253 &  4-09 \\ 
B12-146 &  D10-14 &  13:37:02.07 &  -29:51:58.3 &  32.5 &  0.3 &  - &  6-12 \\ 
B12-147 &  B14-40 &  13:37:02.21 &  -29:49:52.4 &  9.4 &  2.9 &  L14-261 &  3-16 \\ 
B12-148 &  - &  13:37:02.32 &  -29:50:07.0 &  28.2 &  2.5 &  L14-262 &  9-06 \\ 
B12-150 &  B14-41;~D10-15 &  13:37:02.42 &  -29:51:26.1 &  11.2 &  0.8 &  L14-265 &  1-03 \\ 
B14-42 &  - &  13:37:02.89 &  -29:48:39.1 &  25.5 &  4.6 &  - &  9-21 \\ 
B12-151 &  B14-43 &  13:37:03.02 &  -29:49:45.5 &  8.1 &  3.1 &  L14-272 &  3-17 \\ 
B12-154 &  - &  13:37:04.06 &  -29:54:01.9 &  48.3 &  3.2 &  - &  6-10 \\ 
B12-155 &  - &  13:37:04.14 &  -29:53:16.2 &  35.8 &  2.2 &  - &  8-11 \\ 
B12-156 &  - &  13:37:04.41 &  -29:49:38.6 &  13.0 &  3.3 &  L14-287 &  1-02 \\ 
B12-157 &  - &  13:37:04.46 &  -29:53:47.7 &  24.1 &  2.9 &  - &  4-14 \\ 
B12-159 &  - &  13:37:04.50 &  -29:49:35.5 &  16.5 &  3.4 &  L14-288 &  9-07 \\ 
B12-160 &  - &  13:37:04.71 &  -29:55:34.8 &  23.2 &  5.4 &  L14-292 &  7-05 \\ 
B12-162 &  - &  13:37:04.81 &  -29:50:06.8 &  20.9 &  2.7 &  - &  3-18 \\ 
D10-17 &  - &  13:37:04.88 &  -29:52:18.6 &  20.6 &  1.4 &  - &  6-09 \\ 
B14-48 &  - &  13:37:05.44 &  -29:49:18.8 &  9.4 &  3.8 &  - &  9-22 \\ 
B14-49 &  - &  13:37:05.88 &  -29:50:45.5 &  12.6 &  2.1 &  - &  9-23 \\ 
B12-168 &  - &  13:37:06.00 &  -29:50:04.2 &  19.2 &  2.9 &  - &  3-14 \\ 
B12-169 &  - &  13:37:06.04 &  -29:55:14.3 &  12.6 &  5.0 &  L14-310 &  7-11 \\ 
B12-170 &  - &  13:37:06.16 &  -29:54:43.5 &  14.8 &  4.4 &  L14-311 &  5-14 \\ 
B12-171 &  D10-19 &  13:37:06.44 &  -29:50:25.0 &  17.4 &  2.6 &  L14-313 &  9-08 \\ 
B12-172 &  - &  13:37:06.45 &  -29:54:27.3 &  27.7 &  4.1 &  - &  4-03 \\ 
B12-178 &  - &  13:37:07.08 &  -29:53:21.0 &  22.4 &  2.9 &  L14-320 &  7-10 \\ 
B12-179 &  B14-52 &  13:37:07.11 &  -29:51:01.6 &  3.1 &  2.2 &  - &  9-09 \\ 
B12-180 &  D10-22 &  13:37:07.47 &  -29:51:33.4 &  29.2 &  2.0 &  L14-326 &  6-05 \\ 
B12-181 &  - &  13:37:07.50 &  -29:54:16.2 &  24.6 &  4.0 &  - &  8-13 \\ 
B12-183 &  B14-53;~D10-21 &  13:37:07.69 &  -29:51:10.1 &  8.1 &  2.2 &  - &  9-10 \\ 
B12-184 &  - &  13:37:07.71 &  -29:53:01.3 &  17.9 &  2.7 &  - &  4-07 \\ 
B12-185 &  - &  13:37:07.80 &  -29:54:12.2 &  29.5 &  4.0 &  - &  5-16 \\ 
B12-188 &  D10-25 &  13:37:08.10 &  -29:52:21.5 &  17.4 &  2.3 &  - &  8-14 \\ 
B14-54 &  D10-26 &  13:37:08.33 &  -29:50:56.3 &  25.5 &  2.5 &  L14-336 &  9-19 \\ 
B12-191 &  D10-28 &  13:37:08.57 &  -29:51:35.2 &  13.4 &  2.3 &  L14-339 &  6-04 \\ 
B12-193 &  D10-32 &  13:37:08.76 &  -29:51:37.5 &  22.4 &  2.4 &  L14-342 &  3-09 \\ 
B12-194 &  D10-33 &  13:37:09.05 &  -29:51:33.3 &  11.6 &  2.5 &  - &  9-11 \\ 
B12-197 &  D10-36 &  13:37:10.08 &  -29:51:28.2 &  8.9 &  2.8 &  L14-350 &  3-11b \\ 
B12-199 &  D10-37 &  13:37:10.31 &  -29:51:29.0 &  11.6 &  2.9 &  L14-352 &  3-11a \\ 
D10-38 &  - &  13:37:10.36 &  -29:51:33.9 &  25.0 &  2.9 &  - &  6-07 \\ 
B12-201 &  D10-39 &  13:37:10.80 &  -29:51:44.6 &  30.8 &  3.0 &  - &  3-10 \\ 
D10-40 &  - &  13:37:10.84 &  -29:52:44.5 &  8.7 &  3.3 &  - &  8-20 \\ 
B12-202 &  - &  13:37:10.93 &  -29:49:53.0 &  14.8 &  4.0 &  - &  9-12 \\ 
B12-204 &  - &  13:37:11.11 &  -29:53:17.1 &  29.2 &  3.8 &  - &  4-02 \\ 
B12-205 &  - &  13:37:11.34 &  -29:54:19.4 &  43.4 &  4.8 &  - &  5-18 \\ 
B12-207 &  - &  13:37:11.46 &  -29:50:13.6 &  10.3 &  3.8 &  L14-355 &  9-13 \\ 
B12-206 &  - &  13:37:11.47 &  -29:51:41.4 &  13.9 &  3.2 &  L14-356 &  6-01 \\ 
B12-208 &  - &  13:37:11.63 &  -29:51:39.5 &  22.4 &  3.2 &  - &  8-15 \\ 
B12-209 &  - &  13:37:11.88 &  -29:52:15.7 &  24.1 &  3.4 &  - &  4-01 \\ 
B12-210 &  - &  13:37:12.46 &  -29:50:20.1 &  16.1 &  4.0 &  L14-364 &  9-14 \\ 
B12-211 &  - &  13:37:12.81 &  -29:50:12.0 &  15.2 &  4.2 &  L14-368 &  3-06 \\ 
B12-213 &  - &  13:37:13.08 &  -29:51:18.2 &  29.3 &  3.7 &  - &  9-15 \\ 
B12-215 &  - &  13:37:13.97 &  -29:51:51.1 &  32.2 &  4.0 &  - &  9-16 \\ 
B12-219 &  - &  13:37:14.84 &  -29:54:58.6 &  19.2 &  6.3 &  - &  5-19 \\ 
B12-220 &  - &  13:37:16.02 &  -29:53:04.0 &  20.6 &  5.0 &  - &  7-17 \\ 
B12-221 &  B14-62 &  13:37:17.20 &  -29:51:53.4 &  11.2 &  5.0 &  L14-389 &  3-02 \\ 
B12-222 &  - &  13:37:17.27 &  -29:53:25.0 &  25.0 &  5.6 &  - &  7-14 \\ 
B12-223 &  B14-63 &  13:37:17.43 &  -29:51:53.9 &  10.9 &  5.0 &  L14-391 &  9-17 \\ 
\tablenotetext{a}{ Here we have chosen as the primary names, in priority order, numbers from the catalogs of Blair, Winkler \& Long (2012 = B12), Dopita {\it et al.\ }(2010 = D10, Table 2), and Blair {\it et al.\ }(2014 = B14).}
\tablenotetext{b}{ Candidates which are coincident with X-ray sources in M83 as analyzed by Long {\it et al.\ }(2014).}
\tablenotetext{c}{The position and size for B12-014 have been slightly revised since the original B12 paper; see \citet{blair17}.}
\enddata 
\label{table-SNR}
\end{deluxetable}

\begin{deluxetable}{rlccccccl}
\tablecaption{[O III]-selected SNR Candidates With Spectra }
\tablehead{\colhead{Name$^a$} & 
 \colhead{Other~Names} & 
 \colhead{R.A.\,(J2000.)} & 
 \colhead{Decl.\,(J2000.)} & 
 \colhead{Diam.} & 
 \colhead{GC~distance} & 
 \colhead{X-ray~detection$^b$} & 
 \colhead{Mask~ID} 
\\
\colhead{~} & 
 \colhead{~} & 
 \colhead{(h:m:s)} & 
 \colhead{(d:m:s)} & 
 \colhead{(pc)} & 
 \colhead{(kpc)} & 
 \colhead{~} & 
 \colhead{~} 
}
\tabletypesize{\scriptsize}
\tablewidth{0pt}\startdata
B12-304 &  - &  13:36:44.05 &  -29:51:27.1 &  31.3 &  5.2 &  - &  1-18 \\ 
B12-311 &  - &  13:36:52.27 &  -29:54:20.9 &  15.6 &  4.1 &  - &  4-26 \\ 
B12-312 &  - &  13:36:53.60 &  -29:56:00.8 &  13.9 &  6.0 &  L14-110 &  5-02 \\ 
B12-314 &  B14-15 &  13:36:55.27 &  -29:54:02.8 &  5.4 &  3.3 &  L14-135 &  4-16 \\ 
B12-316 &  - &  13:36:58.04 &  -29:49:02.0 &  67.0 &  4.2 &  - &  1-12 \\ 
B12-318 &  - &  13:36:59.03 &  -29:54:58.6 &  22.8 &  4.3 &  - &  7-04a \\ 
B12-320 &  - &  13:36:59.45 &  -29:54:34.7 &  11.6 &  3.7 &  - &  8-16 \\ 
B12-321 &  B14-38 &  13:37:01.27 &  -29:51:59.9 &  2.7 &  0.1 &  L14-243 &  6-11 \\ 
B12-322 &  - &  13:37:02.35 &  -29:54:37.3 &  5.4 &  3.9 &  - &  7-06 \\ 
B12-323 &  - &  13:37:02.38 &  -29:54:15.5 &  8.1 &  3.4 &  - &  4-13 \\ 
B12-324 &  B14-46;~SN1957D &  13:37:03.58 &  -29:49:40.7 &  3.6 &  3.2 &  L14-279 &  1-01 \\ 
D10-4-01 &  - &  13:37:04.12 &  -29:51:03.8 &  7.6 &  1.5 &  - &  3-15 \\ 
B12-326 &  - &  13:37:05.46 &  -29:53:37.2 &  4.9 &  2.9 &  - &  4-05 \\ 
B12-174a &  - &  13:37:06.65 &  -29:53:32.6 &  23.7 &  3.0 &  L14-316 &  6-03 \\ 
D10-4-02 &  - &  13:37:06.97 &  -29:50:57.1 &  13.4 &  2.2 &  - &  3-13 \\ 
B12-333 &  B14-57;~D10-29 &  13:37:08.64 &  -29:52:42.8 &  9.8 &  2.7 &  L14-341 &  8-18 \\ 
B12-336 &  B14-59 &  13:37:12.08 &  -29:50:57.1 &  23.2 &  3.5 &  L14-360 &  3-04 \\ 
B12-339 &  - &  13:37:14.30 &  -29:50:00.8 &  44.7 &  4.7 &  - &  3-07 \\ 
B12-340 &  - &  13:37:14.58 &  -29:50:09.2 &  22.4 &  4.7 &  - &  9-18 \\ 
B12-341 &  - &  13:37:15.21 &  -29:50:39.9 &  18.8 &  4.5 &  - &  3-05 \\ 
B12-343 &  - &  13:37:16.65 &  -29:50:59.9 &  22.4 &  4.8 &  - &  3-08 \\ 
B12-344 &  - &  13:37:17.79 &  -29:51:55.4 &  20.3 &  5.1 &  - &  3-01 \\ 
\tablenotetext{a}{ Here we have chosen as the primary names, in priority order, numbers from the catalogs of Blair, Winkler \& Long (2012 = B12), Dopita {\it et al.\ }(2010 = D10, Table 2), and Blair {\it et al.\ }(2014 = B14).  Sources labeled D10-4-xx refer to Dopita {\it et al.\ }(2010) Table 4. }
\tablenotetext{b}{ Candidates which are coincident with X-ray sources in M83 as analyzed by Long {\it et al.\ }(2014).}
\enddata 
\label{table-O3}
\end{deluxetable}




\begin{deluxetable}{rrrrrrrrrr}
\tablecaption{Emission Line Fluxes: [S II]-selected SNR candidates }
\tablehead{\colhead{Source} & 
 \colhead{H$\alpha$-flux$^a$} & 
 \colhead{H$\beta^b$} & 
 \colhead{[OIII]$\lambda$5007$^b$} & 
 \colhead{[OI]$\lambda$6300$^b$} & 
 \colhead{H$\alpha$} & 
 \colhead{[NII]$\lambda$6584$^b$} & 
 \colhead{[SII]$\lambda$6717$^b$} & 
 \colhead{[SII]$\lambda$6731$^b$} & 
 \colhead{[SII]:H$\alpha$} 
}
\tabletypesize{\scriptsize}
\tablewidth{0pt}\startdata
B12-001 &  29 &  65 &  202 &  55 &  300 &  274 &  167 &  122 &  0.96 \\ 
B12-003 &  52 &  58 &  163 &  69 &  300 &  355 &  120 &  130 &  0.83 \\ 
B12-005 &  12 &  69 &  201 &  67 &  300 &  337 &  142 &  119 &  0.87 \\ 
B12-010 &  63 &  59 &  151 &  27 &  300 &  219 &  106 &  74 &  0.60 \\ 
B12-012 &  425 &  72 &  20 &  13 &  300 &  165 &  92 &  64 &  0.52 \\ 
B12-014 &  11 &  82 &  322 &  34 &  300 &  230 &  92 &  68 &  0.53 \\ 
B12-020 &  264 &  65 &  54 &  14 &  300 &  148 &  87 &  61 &  0.49 \\ 
B12-021 &  590 &  60 &  7 &  7 &  300 &  139 &  62 &  44 &  0.35 \\ 
B12-022 &  384 &  85 &  13 &  18 &  300 &  138 &  72 &  49 &  0.40 \\ 
B12-023 &  381 &  74 &  181 &  40 &  300 &  276 &  154 &  128 &  0.94 \\ 
B12-025 &  169 &  70 &  204 &  63 &  300 &  367 &  212 &  147 &  1.20 \\ 
B12-026 &  7 &  145 &  196 &  -- &  300 &  461 &  215 &  180 &  1.32 \\ 
B12-028 &  253 &  72 &  $\sim$4 &  $\sim$8 &  300 &  117 &  67 &  47 &  0.38 \\ 
B12-031 &  53 &  118 &  224 &  62 &  300 &  384 &  174 &  109 &  0.94 \\ 
B12-033 &  221 &  68 &  41 &  22 &  300 &  164 &  99 &  71 &  0.57 \\ 
B12-034 &  151 &  67 &  61 &  22 &  300 &  199 &  109 &  76 &  0.62 \\ 
B12-036 &  36 &  55 &  137 &  20 &  300 &  161 &  99 &  77 &  0.59 \\ 
B12-035 &  37 &  43 &  -- &  53 &  300 &  196 &  149 &  99 &  0.83 \\ 
B12-037 &  16 &  -- &  65 &  79 &  300 &  436 &  141 &  180 &  1.07 \\ 
B12-039 &  752 &  57 &  42 &  26 &  300 &  237 &  128 &  98 &  0.75 \\ 
B12-043 &  84 &  51 &  $\sim$11 &  18 &  300 &  138 &  95 &  64 &  0.53 \\ 
B12-045 &  418 &  64 &  72 &  23 &  300 &  199 &  79 &  65 &  0.48 \\ 
B12-048 &  278 &  69 &  231 &  10 &  300 &  212 &  109 &  81 &  0.63 \\ 
B14-07 &  233 &  54 &  30 &  8 &  300 &  135 &  44 &  39 &  0.28 \\ 
B14-09 &  1759 &  53 &  9 &  5 &  300 &  125 &  36 &  30 &  0.22 \\ 
B12-058 &  308 &  48 &  48 &  12 &  300 &  153 &  64 &  43 &  0.36 \\ 
B12-057 &  48 &  48 &  152 &  54 &  300 &  342 &  197 &  148 &  1.15 \\ 
B12-065 &  141 &  70 &  90 &  59 &  300 &  231 &  56 &  73 &  0.43 \\ 
B12-067 &  112 &  61 &  144 &  41 &  300 &  293 &  139 &  110 &  0.83 \\ 
B12-066 &  58 &  38 &  45 &  40 &  300 &  275 &  89 &  100 &  0.63 \\ 
B12-069 &  16 &  66 &  182 &  44 &  300 &  308 &  146 &  109 &  0.85 \\ 
B12-073 &  265 &  68 &  76 &  14 &  300 &  160 &  73 &  64 &  0.46 \\ 
B12-074 &  46 &  66 &  296 &  86 &  300 &  571 &  181 &  143 &  1.08 \\ 
B12-077 &  550 &  75 &  14 &  7 &  300 &  110 &  59 &  40 &  0.33 \\ 
B12-084 &  252 &  60 &  204 &  42 &  300 &  283 &  122 &  99 &  0.74 \\ 
B12-087 &  102 &  69 &  241 &  113 &  300 &  630 &  217 &  198 &  1.38 \\ 
B12-089 &  109 &  36 &  98 &  90 &  300 &  563 &  185 &  217 &  1.34 \\ 
B12-091 &  12 &  170 &  250 &  102 &  300 &  392 &  238 &  142 &  1.27 \\ 
B12-097 &  201 &  76 &  72 &  40 &  300 &  296 &  140 &  101 &  0.80 \\ 
B12-098 &  556 &  45 &  88 &  35 &  300 &  289 &  138 &  104 &  0.81 \\ 
B12-101 &  44 &  42 &  340 &  38 &  300 &  450 &  188 &  122 &  1.03 \\ 
B12-104 &  27 &  66 &  208 &  -- &  300 &  230 &  123 &  87 &  0.70 \\ 
B12-106 &  39 &  $\sim$11 &  26 &  137 &  300 &  546 &  116 &  216 &  1.11 \\ 
B12-109 &  50 &  61 &  233 &  57 &  300 &  371 &  176 &  171 &  1.16 \\ 
B12-110 &  32 &  62 &  303 &  63 &  300 &  364 &  190 &  137 &  1.09 \\ 
B12-112 &  88 &  64 &  158 &  51 &  300 &  330 &  179 &  127 &  1.02 \\ 
B12-115 &  213 &  58 &  159 &  78 &  300 &  395 &  67 &  98 &  0.55 \\ 
B12-117 &  1093 &  -- &  60 &  10 &  300 &  139 &  62 &  46 &  0.36 \\ 
B14-19 &  16 &  133 &  296 &  $\sim$36 &  300 &  302 &  131 &  98 &  0.76 \\ 
B12-118 &  129 &  57 &  170 &  17 &  300 &  357 &  105 &  86 &  0.64 \\ 
B14-20 &  26 &  50 &  52 &  117 &  300 &  526 &  259 &  183 &  1.47 \\ 
B12-119 &  137 &  87 &  -- &  -- &  300 &  124 &  60 &  50 &  0.37 \\ 
B12-122 &  140 &  63 &  469 &  49 &  300 &  260 &  112 &  96 &  0.69 \\ 
B12-125 &  76 &  76 &  68 &  48 &  300 &  251 &  160 &  110 &  0.90 \\ 
B12-124 &  73 &  55 &  273 &  107 &  300 &  740 &  245 &  241 &  1.62 \\ 
B12-127 &  157 &  76 &  240 &  43 &  300 &  338 &  95 &  119 &  0.71 \\ 
B12-129 &  13 &  65 &  67 &  41 &  300 &  241 &  70 &  71 &  0.47 \\ 
B12-130 &  107 &  60 &  62 &  28 &  300 &  209 &  98 &  66 &  0.55 \\ 
B12-132 &  37 &  49 &  379 &  64 &  300 &  551 &  242 &  174 &  1.39 \\ 
B12-133 &  8 &  58 &  216 &  -- &  300 &  493 &  258 &  213 &  1.57 \\ 
B12-134 &  325 &  45 &  26 &  23 &  300 &  171 &  91 &  64 &  0.52 \\ 
B12-136 &  253 &  68 &  -- &  $\sim$3 &  300 &  123 &  63 &  45 &  0.36 \\ 
B12-137 &  33 &  55 &  442 &  45 &  300 &  479 &  190 &  178 &  1.23 \\ 
B12-142 &  104 &  76 &  222 &  39 &  300 &  330 &  162 &  126 &  0.96 \\ 
B12-146 &  118 &  69 &  78 &  131 &  300 &  450 &  253 &  187 &  1.47 \\ 
B12-147 &  112 &  30 &  93 &  85 &  300 &  437 &  158 &  177 &  1.12 \\ 
B12-148 &  82 &  84 &  230 &  -- &  300 &  224 &  114 &  78 &  0.64 \\ 
B12-150 &  198 &  59 &  172 &  62 &  300 &  491 &  149 &  160 &  1.03 \\ 
B14-42 &  1275 &  108 &  19 &  4 &  300 &  136 &  54 &  39 &  0.31 \\ 
B12-151 &  457 &  71 &  19 &  26 &  300 &  203 &  71 &  73 &  0.48 \\ 
B12-154 &  85 &  72 &  302 &  88 &  300 &  500 &  308 &  238 &  1.82 \\ 
B12-155 &  44 &  84 &  310 &  $\sim$217 &  300 &  450 &  166 &  125 &  0.97 \\ 
B12-156 &  330 &  80 &  132 &  22 &  300 &  204 &  96 &  87 &  0.61 \\ 
B12-157 &  241 &  63 &  102 &  15 &  300 &  174 &  83 &  58 &  0.47 \\ 
B12-159 &  257 &  63 &  174 &  53 &  300 &  288 &  163 &  123 &  0.95 \\ 
B12-160 &  84 &  60 &  429 &  120 &  300 &  658 &  278 &  217 &  1.65 \\ 
B12-162 &  40 &  62 &  209 &  87 &  300 &  607 &  328 &  239 &  1.89 \\ 
D10-17 &  23 &  57 &  80 &  90 &  300 &  410 &  201 &  131 &  1.11 \\ 
B14-48 &  562 &  46 &  37 &  $\sim$4 &  300 &  127 &  24 &  15 &  0.13 \\ 
B14-49 &  643 &  58 &  28 &  -- &  300 &  103 &  33 &  30 &  0.21 \\ 
B12-168 &  42 &  62 &  242 &  103 &  300 &  504 &  265 &  194 &  1.53 \\ 
B12-169 &  165 &  61 &  173 &  83 &  300 &  483 &  139 &  157 &  0.99 \\ 
B12-170 &  27 &  68 &  529 &  30 &  300 &  319 &  144 &  118 &  0.87 \\ 
B12-171 &  251 &  83 &  74 &  59 &  300 &  305 &  144 &  113 &  0.86 \\ 
B12-172 &  78 &  64 &  308 &  35 &  300 &  344 &  166 &  120 &  0.95 \\ 
B12-178 &  166 &  80 &  85 &  45 &  300 &  255 &  137 &  102 &  0.80 \\ 
B12-179 &  101 &  62 &  50 &  40 &  300 &  228 &  70 &  75 &  0.49 \\ 
B12-180 &  264 &  58 &  138 &  80 &  300 &  424 &  204 &  151 &  1.18 \\ 
B12-181 &  39 &  86 &  141 &  114 &  300 &  430 &  245 &  179 &  1.41 \\ 
B12-183 &  652 &  53 &  27 &  14 &  300 &  136 &  50 &  38 &  0.29 \\ 
B12-184 &  106 &  58 &  64 &  16 &  300 &  222 &  99 &  72 &  0.57 \\ 
B12-185 &  115 &  55 &  31 &  28 &  300 &  192 &  120 &  82 &  0.67 \\ 
B12-188 &  49 &  96 &  $\sim$49 &  $\sim$42 &  300 &  252 &  136 &  83 &  0.73 \\ 
B14-54 &  306 &  91 &  92 &  22 &  300 &  191 &  109 &  81 &  0.63 \\ 
B12-191 &  89 &  49 &  165 &  73 &  300 &  446 &  201 &  178 &  1.26 \\ 
B12-193 &  120 &  35 &  63 &  46 &  300 &  343 &  143 &  122 &  0.88 \\ 
B12-194 &  32 &  -- &  $\sim$120 &  -- &  300 &  340 &  258 &  208 &  1.55 \\ 
B12-197 &  916 &  50 &  14 &  10 &  300 &  157 &  58 &  48 &  0.35 \\ 
B12-199 &  469 &  74 &  58 &  11 &  300 &  148 &  72 &  52 &  0.41 \\ 
D10-38 &  72 &  62 &  111 &  $\sim$13 &  300 &  177 &  109 &  81 &  0.63 \\ 
B12-201 &  40 &  86 &  343 &  51 &  300 &  414 &  189 &  145 &  1.11 \\ 
D10-40 &  14 &  74 &  139 &  $\sim$300 &  300 &  238 &  193 &  180 &  1.24 \\ 
B12-202 &  239 &  71 &  41 &  $\sim$22 &  300 &  161 &  93 &  70 &  0.54 \\ 
B12-204 &  16 &  79 &  599 &  $\sim$49 &  300 &  489 &  229 &  185 &  1.38 \\ 
B12-205 &  221 &  56 &  13 &  19 &  300 &  124 &  72 &  53 &  0.42 \\ 
B12-207 &  43 &  $\sim$90 &  300 &  $\sim$132 &  300 &  482 &  130 &  123 &  0.84 \\ 
B12-206 &  80 &  61 &  216 &  71 &  300 &  447 &  211 &  166 &  1.25 \\ 
B12-208 &  76 &  68 &  128 &  36 &  300 &  237 &  101 &  73 &  0.58 \\ 
B12-209 &  297 &  81 &  183 &  28 &  300 &  237 &  89 &  84 &  0.58 \\ 
B12-210 &  1082 &  80 &  13 &  18 &  300 &  142 &  57 &  46 &  0.34 \\ 
B12-211 &  41 &  66 &  203 &  -- &  300 &  211 &  106 &  76 &  0.61 \\ 
B12-213 &  26 &  165 &  250 &  129 &  300 &  503 &  353 &  261 &  2.05 \\ 
B12-215 &  63 &  56 &  366 &  -- &  300 &  304 &  151 &  97 &  0.83 \\ 
B12-219 &  189 &  59 &  107 &  29 &  300 &  279 &  158 &  117 &  0.91 \\ 
B12-220 &  20 &  39 &  417 &  68 &  300 &  596 &  182 &  143 &  1.08 \\ 
B12-221 &  236 &  42 &  150 &  64 &  300 &  378 &  111 &  143 &  0.85 \\ 
B12-222 &  88 &  66 &  176 &  40 &  300 &  334 &  196 &  147 &  1.14 \\ 
B12-223 &  602 &  57 &  41 &  44 &  300 &  208 &  90 &  81 &  0.57 \\ 
\tablenotetext{a}{ H$\alpha $ flux in units of 10$^{-17}$ ergs s$^{-1}$}
\tablenotetext{b}{ Ratio to H$\alpha $ flux where, by convention, H$\alpha$ is normalized to 300.}
\enddata 
\label{table-snr-flux}
\end{deluxetable}


\begin{deluxetable}{rrrrrrrrrr}
\tablecaption{Emission Line Fluxes: [O III]-selected SNR candidates }
\tablehead{\colhead{Source} & 
 \colhead{H$\alpha$-flux$^a$} & 
 \colhead{H$\beta^b$} & 
 \colhead{[OIII]$\lambda$5007$^b$} & 
 \colhead{[OI]$\lambda$6300$^b$} & 
 \colhead{H$\alpha$} & 
 \colhead{[NII]$\lambda$6584$^b$} & 
 \colhead{[SII]$\lambda$6717$^b$} & 
 \colhead{[SII]$\lambda$6731$^b$} & 
 \colhead{[SII]:H$\alpha$} 
}
\tabletypesize{\scriptsize}
\tablewidth{0pt}\startdata
B12-304 &  135 &  56 &  76 &  7 &  300 &  109 &  37 &  25 &  0.21 \\ 
B12-311 &  82 &  59 &  197 &  $\sim$7 &  300 &  172 &  48 &  34 &  0.27 \\ 
B12-312 &  14 &  54 &  177 &  $\sim$39 &  300 &  234 &  65 &  43 &  0.36 \\ 
B12-314 &  12 &  77 &  855 &  -- &  300 &  90 &  -- &  -- &  -- \\ 
B12-316 &  131 &  54 &  107 &  -- &  300 &  125 &  45 &  32 &  0.26 \\ 
B12-318 &  120 &  82 &  135 &  -- &  300 &  92 &  33 &  24 &  0.19 \\ 
B12-320 &  126 &  101 &  491 &  21 &  300 &  166 &  52 &  36 &  0.29 \\ 
B12-321 &  302 &  57 &  41 &  9 &  300 &  184 &  53 &  46 &  0.33 \\ 
B12-322 &  115 &  82 &  126 &  -- &  300 &  118 &  26 &  19 &  0.15 \\ 
B12-323 &  206 &  72 &  60 &  $\sim$3 &  300 &  125 &  34 &  23 &  0.19 \\ 
B12-324 &  69 &  49 &  $\sim$78 &  -- &  300 &  152 &  39 &  29 &  0.23 \\ 
D10-1-01 &  137 &  49 &  115 &  18 &  300 &  203 &  60 &  40 &  0.33 \\ 
B12-326 &  142 &  61 &  126 &  6 &  300 &  124 &  40 &  28 &  0.22 \\ 
B12-174a$^c$ &  -- &  -- &  -- &  -- &  300 &  -- &  -- &  -- &  -- \\ 
D10-1-02 &  104 &  73 &  120 &  $\sim$9 &  300 &  144 &  47 &  29 &  0.25 \\ 
B12-333 &  982 &  100 &  21 &  $\sim$3 &  300 &  113 &  35 &  25 &  0.20 \\ 
B12-336 &  419 &  58 &  29 &  -- &  300 &  123 &  23 &  18 &  0.14 \\ 
B12-339 &  49 &  52 &  470 &  -- &  300 &  134 &  49 &  36 &  0.28 \\ 
B12-340 &  332 &  88 &  196 &  17 &  300 &  148 &  38 &  28 &  0.22 \\ 
B12-341 &  168 &  54 &  149 &  -- &  300 &  111 &  44 &  30 &  0.25 \\ 
B12-343 &  111 &  92 &  154 &  -- &  300 &  119 &  45 &  31 &  0.25 \\ 
B12-344 &  86 &  66 &  395 &  -- &  300 &  82 &  33 &  25 &  0.19 \\ 
\tablenotetext{a}{ H$\alpha $ flux in units of 10$^{-17} \FLUX$\@.  This is flux within the slit, not necessarily the entire flux from the object.}
\tablenotetext{b}{ Ratio to H$\alpha $ flux where, by convention, H$\alpha$ is normalized to 300.}
\tablenotetext{c}{ B12-174a, which has broad,  blended emission lines, is discussed by Blair {\it et al.\ }(2015).}
\enddata 
\label{table-O3-flux}
\end{deluxetable}


\begin{deluxetable}{rcccr}
\tablecaption{H II Regions with Spectra }
\tablehead{\colhead{Name} & 
 \colhead{R.A.\,(J2000.)} & 
 \colhead{Decl.\,(J2000.)} & 
 \colhead{GC~distance} & 
 \colhead{Mask~ID} 
\\
\colhead{~} & 
 \colhead{(h:m:s)} & 
 \colhead{(d:m:s)} & 
 \colhead{(kpc)} & 
 \colhead{~} 
}
\tabletypesize{\scriptsize}
\tablewidth{0pt}\startdata
HII-01 &  13:36:41.01 &  -29:51:57.0 &  6.1 &  2-03 \\ 
HII-02 &  13:36:42.21 &  -29:52:31.8 &  5.7 &  2-05 \\ 
HII-03 &  13:36:42.38 &  -29:51:18.0 &  5.8 &  1-22 \\ 
HII-04 &  13:36:43.43 &  -29:52:23.6 &  5.3 &  2-04 \\ 
HII-05 &  13:36:44.06 &  -29:51:27.0 &  5.2 &  1-18a \\ 
HII-06 &  13:36:45.23 &  -29:49:23.9 &  6.2 &  1-20 \\ 
HII-07 &  13:36:46.29 &  -29:53:43.3 &  4.9 &  2-07b \\ 
HII-08 &  13:36:54.73 &  -29:52:57.0 &  2.3 &  4-20 \\ 
HII-09 &  13:36:56.86 &  -29:52:48.7 &  1.7 &  2-18 \\ 
HII-10 &  13:36:59.26 &  -29:54:58.3 &  4.3 &  7-04b \\
HII-11 &  13:36:59.39 &  -29:54:58.5 &  4.3 &  7-04c \\
HII-12 &  13:37:00.03 &  -29:52:19.3 &  0.6 &  6-18 \\ 
HII-13 &  13:37:00.67 &  -29:54:26.6 &  3.5 &  4-11a \\ 
HII-14 &  13:37:03.40 &  -29:54:02.3 &  3.1 &  4-15 \\ 
HII-15 &  13:37:07.72 &  -29:53:01.4 &  2.7 &  4-07a \\ 
HII-16 &  13:37:09.48 &  -29:49:25.6 &  4.2 &  3-12 \\ 
HII-17 &  13:37:09.75 &  -29:52:44.0 &  3.0 &  4-08 \\ 
HII-18 &  13:37:11.39 &  -29:55:35.1 &  6.3 &  7-16 \\ 
\enddata 
\label{h2_regions}
\end{deluxetable}


\begin{deluxetable}{rrrrrrrrrr}
\tablecaption{Emission Line Fluxes: HII Regions }
\tablehead{\colhead{Source} & 
 \colhead{H$\alpha$-flux$^a$} & 
 \colhead{H$\beta^b$} & 
 \colhead{[OIII]$\lambda$5007$^b$} & 
 \colhead{[OI]$\lambda$6300$^b$} & 
 \colhead{H$\alpha$} & 
 \colhead{[NII]$\lambda$6584$^b$} & 
 \colhead{[SII]$\lambda$6717$^b$} & 
 \colhead{[SII]$\lambda$6731$^b$} & 
 \colhead{[SII]:H$\alpha$} 
}
\tabletypesize{\scriptsize}
\tablewidth{0pt}\startdata
HII-01 &  1348 &  71 &  25 &  3 &  300 &  127 &  49 &  37 &  0.29 \\ 
HII-02 &  1930 &  43 &  18 &  $\sim$1 &  300 &  101 &  25 &  18 &  0.14 \\ 
HII-03 &  902 &  62 &  52 &  $\sim$2 &  300 &  119 &  23 &  18 &  0.14 \\ 
HII-04 &  9809 &  62 &  31 &  1 &  300 &  114 &  25 &  18 &  0.14 \\ 
HII-05 &  396 &  54 &  6 &  $\sim$2 &  300 &  102 &  30 &  20 &  0.17 \\ 
HII-06 &  732 &  56 &  5 &  $\sim$2 &  300 &  102 &  25 &  18 &  0.14 \\ 
HII-07 &  90 &  70 &  13 &  -- &  300 &  129 &  33 &  23 &  0.19 \\ 
HII-08 &  2073 &  34 &  1 &  1 &  300 &  108 &  34 &  26 &  0.20 \\ 
HII-09 &  2848 &  34 &  4 &  $\sim$1 &  300 &  131 &  26 &  20 &  0.16 \\ 
HII-10 &  266 &  78 &  31 &  -- &  300 &  106 &  37 &  25 &  0.21 \\ 
HII-11 &  133 &  96 &  79 &  -- &  300 &  138 &  50 &  33 &  0.28 \\ 
HII-12 &  4649 &  40 &  7 &  1 &  300 &  97 &  30 &  25 &  0.18 \\ 
HII-13 &  657 &  43 &  9 &  4 &  300 &  133 &  49 &  35 &  0.28 \\ 
HII-14 &  7206 &  50 &  16 &  2 &  300 &  154 &  31 &  27 &  0.19 \\ 
HII-15 &  341 &  55 &  29 &  $\sim$2 &  300 &  125 &  25 &  18 &  0.14 \\ 
HII-16 &  1040 &  60 &  27 &  -- &  300 &  131 &  22 &  16 &  0.13 \\ 
HII-17 &  1023 &  67 &  34 &  3 &  300 &  106 &  29 &  20 &  0.16 \\ 
HII-18 &  2331 &  63 &  14 &  -- &  300 &  123 &  23 &  17 &  0.13 \\ 
\tablenotetext{a}{ H$\alpha $ flux in units of 10$^{-17}$ ergs s$^{-1}$}
\tablenotetext{b}{ Ratio to H$\alpha $ flux where, by convention, H$\alpha$ is normalized to 300.}
\enddata 
\label{table-HII-flux}
\end{deluxetable}


\begin{deluxetable}{lcccccrcl}
\tabletypesize{\footnotesize}
\rotate
\tablewidth{0pt}
\tablecaption{Properties of Known Ejecta-Dominated Supernova Remnants}

\tablehead{
\colhead {Object}  & 
\colhead {Distance} &  
\colhead {Diameter}  &  
\colhead{Age}  &
\colhead{$A_{\rm V}$} &
\colhead{$F_{{\rm [O\,III]}}$\tablenotemark{a}}  &  
\colhead{$F_{{\rm X}\,(0.3-2.1\,{\rm keV})}$\tablenotemark{a}}  & 
\colhead{$F_{r}$\,(1 GHz)}  & 
\colhead{References}  \\ 

\colhead {}  & 
\colhead {(kpc)} &  
\colhead {(pc)}  &  
\colhead{(yr)}  & 
\colhead{(mag)} &
\colhead{{\scriptsize($10^{-17}\FLUX$)}}  &  
\colhead{{\scriptsize($10^{-16}\FLUX$)}}  & 
\colhead{(mJy)}   &
\colhead{}   
}

\startdata
		Cas A & 3.4 & 5 &  $\sim 335$ & 6.2\phn & 91\tablenotemark{b} & 102\phn\phn\phn\phn\phn\phn\phn & 1.48 & 1, 2, 3, 4, 5, 6  \\
		G292.0+1.8 & 6 &16 & $\sim 3000$ & 3.3\phn &20\tablenotemark{b}& 35\phn\phn\phn\phn\phn\phn\phn & 0.03 & 5, 6, 7, 8, 9, 10  \\
		Puppis A & 2.2 &32 & $\sim 3500$ & & c & 98\phn\phn\phn\phn\phn\phn\phn & 0.03 & 6, 11, 12, 13  \\
		N132D (LMC) & 50 &25 & $\sim 3150$ & 0.62 & 60\tablenotemark{b,d} &200\phn\phn\phn\phn\phn\phn\phn &0.62 &  4, 5, 14, 15, 16, 17  \\
		0540--69.3 (LMC) & 50  &2 & 760 - 1660 & 0.59 & 1 & 12\phn\phn\phn\phn\phn\phn\phn & 0.12 &  5, 14, 15, 17, 18 \\
		E0102--72.3 (SMC) & 60 &13 & $\sim 2050$ & 0.25 & 39 & 100\phn\phn\phn\phn\phn\phn\phn & 0.06 &  5, 14, 19,  20, 21\\
		SNR NGC\,4449-1  & 3820 &$\sim 0.7\phn\phn$ & $\sim 55$ & 0.7\phn & 13500\phn\phn\phn & 900\phn\phn\phn\phn\phn\phn\phn & 5.8\phn & 22, 23, 24, 25, 26  \\
		SN1957D  &  4610  &  & 59 & & 42\tablenotemark{e} & 67\phn\phn\phn\phn\phn\phn\phn & 0.50 & 27 \\ [2 pt]
\hline\\[-11pt]

		Crab Nebula\tablenotemark{f} & 2 &  4 & 962 & 2.2 & 33 & 30\phn\phn\phn\phn\phn\phn\phn & 0.20 & 5, 6, 28, 29  \\
\enddata

\tablerefs{
(1) \citet{reed95}; (2) \citet{thorstensen01}; (3) \citet{eriksen09}; (4) unpublished data from Burrell or  Curtis Schmidt images obtained as part of an early feasibility study for extragalactic O-rich SNR searches; (5) Chandra SNR Catalog, http://hea-www.cfa.harvard.edu/ChandraSNR/;  (6) \citet{green14} (7) \citet{gaensler03}; (8) \citet{winkler09}; (9) \citet{lee10}; (10) \citet{winkler06};  (11) \citet{reynoso03}; (12) \citet{winkler88}; (13) ROSAT data, \citet{petre96} and PIMMS: http://cxc.harvard.edu/toolkit/pimms.jsp; (14) \citet{vandenbergh89}; (15) \citet{morse95}; (16) \citet{filipovic98}; (17) \citet{morse06}; (18) \citet{brantseg14};
(19) \citet{finkelstein06}; (20) \citet{blair89}; 
(21) \citet{payne04}; (22) \citet{annibali08}; (23) \citet{mezcua13}; (24) \citet{milisavljevic08};  (25) \citet{patnaude03};  (26) \citet{lacey07}; (27) \citet{long12} (28) \citet{trimble73}; (29) \citet{smith03}.    
}
\tablenotetext{a}{Measured \oiii\ fluxes from each object have been corrected for absorption, and  converted to a distance of 4.61 Mpc.}
\tablenotetext{b}{\oiii\ flux is measured in images with a 45 \AA\ (FWHM) bandpass.}
\tablenotetext{c}{The vast majority of the optical flux from Puppis A stems from radiative filaments of shocked CSM\@. It is highly unlikely that the small fraction of broad-line emission from ejecta knots would have been detected from a Puppis A analog in M83.}
\tablenotetext{d}{N132D \oiii\ flux is for the inner ring of fast filaments only; including the outer ``horseshoe" of narrow-line emission would give an unabsorbed \oiii\ flux of $156 \times 10^{-17}\FLUX$ (of which about one third would be broad-line) for an N132D analog in M83.}
\tablenotetext{e}{Broad-line components only.}
\tablenotetext{f}{Crab Nebula included for comparison purposes only.}

\label{table_osnrs}
\end{deluxetable}

\clearpage

\end{document}